\documentclass[12pt,a4paper]{article}
\pdfoutput=1
\usepackage{bm, amssymb,pifont,cancel, amsmath,comment,color}
\usepackage{here}
\usepackage{cite}
\usepackage{graphicx}
\usepackage{subfigure}

\makeatletter

\newcommand{\be}{\begin{equation}}
\newcommand{\ee}{\end{equation}}
\newcommand{\bea}{\begin{eqnarray}}
\newcommand{\eea}{\end{eqnarray}}

\@addtoreset{equation}{section}
\makeatother

\begin{document}

\begin{titlepage}

\begin{centering}
\vspace{1cm}
{\Large {\bf Inflation and Supersymmetry Breaking  \vspace{.1cm} \\ in Higgs-$R^2$ Supergravity }} \\

\vspace{1.5cm}

{\bf  Shuntaro Aoki$^{\dagger}$, Hyun Min Lee$^{*}$, and Adriana G. Menkara$^\ddagger$}

\vspace{.5cm}

{\it  Department of Physics, Chung-Ang University, Seoul 06974, Korea.}

\end{centering}

\vspace{2cm}

\begin{abstract}
\noindent
We propose a new construction of the supergravity inflation as an UV completion of the Higgs-$R^2$ inflation. 
In the dual description of $R^2$-supergravity, we show that there appear dual chiral superfields containing the scalaron or sigma field in the Starobinsky inflation, which unitarizes the supersymmetric Higgs inflation with a large non-minimal coupling up to the Planck scale.  We find that a successful slow-roll inflation is achievable in the Higgs-sigma field space, but under the condition that higher curvature terms are introduced to cure the tachyonic mass problems for spectator singlet scalar fields. We also discuss supersymmetry breaking and its transmission to the visible sector as a result of the couplings of the dual chiral superfields and the non-minimal gravity coupling of the Higgs fields.

\end{abstract}

\vspace{3cm}

\begin{flushleft} 
$^\dagger$Email:  shuntaro@cau.ac.kr  \\
$^*$Email: hminlee@cau.ac.kr \\
$^\ddagger$Email: amenkara@cau.ac.kr
\end{flushleft}

\end{titlepage}

\tableofcontents
\vspace{35pt}
\hrule

\section{Introduction}\label{intro}

The observed Cosmic Microwave Background (CMB) not only strongly supports the existence of inflation in the early universe, but also constrains many inflation models via curvature and tensor perturbations~\cite{Planck:2018jri}. The Higgs inflation~\cite{Bezrukov:2007ep} and the Starobinsky ($R^2$-) inflation models~\cite{Starobinsky:1980te,Mukhanov:1981xt,Starobinsky:1983zz} have attracted a lot of attention as successful minimal extensions of the Standard Model (SM) for inflation. In the Higgs inflation, the SM Higgs boson  plays the role of inflaton with a non-minimal coupling to the Ricci scalar. In the Starobinsky inflation, the general relativity is modified with an $R^2$ term, which gives rise to a dual scalar field (scalaron) as the inflaton. Both of these inflation models predict similar types of scalar potential, in perfect agreement with the observed data.

The unitarity problem occurs due to the large non-minimal coupling in the original Higgs inflation \cite{Burgess:2009ea,Barbon:2009ya,Burgess:2010zq,Hertzberg:2010dc}, so there is a need of introducing an extra degree of freedom below the unitarity scale \cite{Giudice:2010ka,Elias-Miro:2012eoi,Lee:2013nv,largeVEV,linear1,linear2}. 
In this regard, the possibility of combining the Higgs and Starobinsky  models has been revisited due to the presence of the dual scalar field, scalaron, in the Starobinsky model~\cite{add,Ema:2017rqn,Gorbunov:2018llf,add2}.  Indeed, it has been shown that the scalaron or sigma field in the linearized action, in the Starobinsky model,  unitarizes the Higgs inflation up to the Planck scale~\cite{He:2018mgb,Cheong:2020rao,Ema:2020zvg,Ema:2020evi,Lee:2021dgi}. There are more general sigma models unitarizing the Higgs inflation \cite{Lee:2021dgi} beyond the Starobinsky model.

In this article, we construct a Next-to-Minimal Supersymmetric Standard Model (NMSSM) extension of the Higgs inflation in $R^2$-supergravity and its dual-scalar supergravity framework. There are a variety of motivations for low-energy supersymmetry (SUSY) such as solutions to the hierarchy problem, gauge coupling unification, vacuum instability problem, natural candidates for dark matter, etc. However, there has been no convincing evidence for supersymmetric particles at the Large Hadron Collider (LHC) or precision measurements, so there might be a little hierarchy between the weak scale and the superparticle masses. Nonetheless, for the consistency of non-supersymmetric models at high energies and the quantum theory of gravity, it is necessary to develop supergravity extensions of the inflation. 

As it comes to a supersymmetric extension of inflation models, there is a question on the influences of extra scalar fields on the inflatonary trajectory. For instance, the Minimal Supersymmetric Standard Model (MSSM) for the Higgs inflation has been studied in the framework of Jordan frame supergravity. In this case, the Higgs potential stems from D-terms, so it vanishes along the D-flat direction for which the second Higgs field is stabilized~{\cite{Einhorn:2009bh}. As a result, the model has been extended to the NMSSM where a singlet chiral multiplet $S$ provides an additional Higgs potential. 
Even in this case, however, the singlet scalar $S$ becomes tachyonic during inflation, destablizing the Higgs inflation~\cite{Ferrara:2010yw}. Therefore, it is necessary to extend the minimal frame function by a quartic term for the singlet scalar $S$~\cite{Lee:2010hj,Ferrara:2010in} (See also Refs.~\cite{Yamaguchi:2000vm,Kitano:2006wz,Kallosh:2006dv}). The lesson here is that it is important to check the consistency in every extension of the minimal inflation models. A similar discussion is also applied for the supergravity extension of the Starobinaky inflation~\cite{Cecotti:1987sa,Ketov:2010qz,Ketov:2012yz,Ketov:2012jt,Kallosh:2013lkr,Farakos:2013cqa,Ferrara:2013wka}: additional scalar fields required for supersymmetry lead to a  tachyonic instability and would destabilize the inflation~\cite{Kallosh:2013lkr}. Therefore, we also study these stability issues in the $R^2$-supergravity extensions of the Higgs-sigma field inflation.

We also discuss the importance of the equivalent frames describing the same physics, because some symmetries and critical problems can be made more clear in one frame than in the other. In the sigma-model frame for Higgs inflation where the conformal symmetry is manifest, the Higgs kinetic term can be recast into a non-linear model type \cite{Ema:2020zvg,Lee:2021dgi}, so the unitarity problem can be seen clearly from the non-canonical form of the Higgs kinetic terms \cite{Burgess:2010zq,Giudice:2010ka}.  Moreover, in the presence of the $R^2$ term in the Higgs inflation, the dual-scalar field for the $R^2$ term, called the sigma field, appears to respect the conformal symmetry and internal symmetries in the sigma-model frame, so unitarity becomes manifest \cite{Ema:2020zvg,Lee:2021dgi}.
We introduce the equivalent frames  for the Higgs-sigma inflation in a manifestly supersymmetric way in conformal supergravity~\cite{Kaku:1978nz,Kaku:1978ea,Townsend:1979ki,Kugo:1982cu,Kugo:1983mv}, thanks to a large gauge symmetry.

In our setup, we pursue a comprehensive description of the early Universe in a supergravity inflation model, from the inflationary dynamics towards the low-energy phenomenology after SUSY breaking in the vacuum. We show the important roles of dual chiral superfields  appearing in the dual description of $R^2$-supergravity, such as the UV completion of the supersymmetric Higgs inflation with a large non-minimal coupling as well as the stability of the slow-roll inflation. We also address  the effects of the extra chiral multiplets for SUSY breaking and its mediation to the visible sector.

The paper is organized as follows. First, in Sec.~\ref{S1}, we introduce the supersymmetric generalization of the Higgs and Starobinsky models where both a non-minimal gravity coupling for the NMSSM Higgs fields and an $R^2$ term are explicitly introduced in $R^2$-supergravity. Next, we derive the dual-scalar Lagrangian at the supergravity level where the supersymmetric $R^2$ term is converted to extra singlet chiral multiplets. Then, in Sec.~\ref{S2}, we present the bosonic Lagrangians for NMSSM in the dual-scalar supergravity, in equivalent frames, such as Jordan, Einstein and sigma-model frames. In Sec~\ref{S3}, we continue to study the effective action for inflation in our model and check the stability of heavy scalars for consistency.  In Sec~\ref{Pheno}, we consider the mechanisms for SUSY breaking in the presence of higher curvature terms or an extra singlet chiral multiplet, and discuss  the mediation of SUSY breaking to the visible sector.
Finally, conclusions are drawn.


\section{$R^2$-supergravity and dual description}\label{S1}

We provide the model setup for the NMSSM Higgs inflation in $R^2$-supergravity with superconformal  symmetry and consider the dual-scalar description of $R^2$-supergravity in terms of two singlet chiral superfields, $T$ and $C$.

\subsection{$R^2$-supergravity and NMSSM}

In order to include the $R^2$ term and the non-minimal coupling for Higgs fields in 4D supergravity, we  consider the action in superconformal setup~\cite{Kaku:1978nz,Kaku:1978ea,Townsend:1979ki,Kugo:1982cu,Kugo:1983mv},\footnote{We mostly follow the conventions of Ref.~\cite{Freedman:2012zz}.}  as follows, 
\begin{align}
S=[|X^0|^2\tilde{\Omega}(z^{\alpha},\bar{z}^{\bar{\beta}})]_D+[(X^0)^3\tilde{W}(z^{\alpha})]_F+[f_{AB}(z^{\alpha})\bar{\mathcal{W}}^A\mathcal{W}^B]_F+[\alpha \bar{\mathcal{R}}\mathcal{R}]_D,   \label{S_HD} 
\end{align}
where $[...]_{D,F}$ denote the superconformal $D$- and $F$-term formulae, which are applicable for real and chiral multiplets with (Weyl weight, chiral weight)$=(2,0)$ and $(3,3)$, respectively. Here, $X^0$ is a chiral compensator multiplet with $(1,1)$, and $\bar{X}^{\bar{0}}$ is its conjugate with $(1,-1)$.
$z^{\alpha}$ and $\bar{z}^{\bar{\beta}}$ are chiral and anti-chiral matter multiplets with $(0,0)$, which will be identified as NMSSM chiral superfields later. $\tilde{\Omega} $ is a real function of matter multiplets, called the frame function, and it is related to the K\"ahler potential by $\tilde{\mathcal{K}}=-3\log \left(-\frac{\tilde{\Omega}}{3} \right)$. $\tilde{W}$ and $f$ are the superpotential and the gauge kinetic function, respectively, which are the holomorphic functions of $z^{\alpha}$. $\mathcal{W}^A$ denotes a gauge field-strength multiplet with $A$ being the gauge indices.  

The last term of Eq.~$\eqref{S_HD}$ contains a curvature multiplet $\mathcal{R}$~\cite{Kugo:1983mv} with weights $(1,1)$, which is defined as 
\begin{align}
\mathcal{R}=(X^0)^{-1} \Sigma (\bar{X}^{\bar{0}}),
\end{align}
where $\Sigma $ is a chiral projection operator. The coefficient of the last term, $\alpha$, is a real parameter and is taken to be positive for stability reason. When $\alpha=0$, the action $\eqref{S_HD} $ is reduced to the standard superconformal action up to second derivatives. In our case, we take a nonzero $\alpha$ for which the $R^2$ term and some new dynamical degree of freedoms are included~\cite{Cecotti:1987sa}. 

In the superconformal construction, the compensator multiplet $X^0$ is an unphysical degree of freedom which would be eliminated by the superconformal gauge fixing conditions. We impose the dilatation gauge condition by $X^0=1$ on the lowest scalar component of $X^0$-multiplet,\footnote{We sometimes use the same characters for superfields and their lowest scalar components.} in order to obtain the action with the field-dependent Einstein term from the product of  $\tilde{\Omega}$ and $R$ (i.e., Jordan frame action).   
Then, after integrating out some auxiliary fields, we obtain the following bosonic part of the Lagrangian:
\begin{align}
\nonumber\mathcal{L}/\sqrt{-g}=&-\tilde{\Omega}_{\alpha\bar{\beta}}\partial_{\mu} z^{\alpha} \partial^{\mu} \bar{z}^{\bar{\beta}}+(-i\tilde{\Omega}_{\alpha}\partial_{\mu} z^{\alpha} \mathcal{A}^{\mu}+{\rm{c.c.}})+\tilde{\Omega}(-\mathcal{A}^{2}+\left|F^{0}\right|^{2})+(3F^0\tilde{W}+{\rm{c.c.}})\\
\nonumber&+\left(-\frac{\tilde{\Omega}}{6}+\frac{\alpha}{6}|F^0|^2+\frac{\alpha}{3}\mathcal{A}^2\right)R+\frac{\alpha}{36}R^2+\alpha\left(\mathcal{A}^{2}+\left|F^{0}\right|^{2}\right)^{2}+\alpha(\nabla_{\mu}  \mathcal{A}^{\mu})^{2}\\
&-\alpha\left|\partial_{\mu} F^{0}-3 i \mathcal{A}_{\mu} F^{0}\right|^{2}-\tilde{\Omega}^{\alpha\bar{\beta}}(\tilde{\Omega}_{\alpha}\bar{F}^{\bar{0}}+\tilde{W}_{\alpha})(\tilde{\Omega}_{\bar{\beta}}F^{0}+\bar{\tilde{W}}_{\bar{\beta}})\nonumber \\ 
&-\frac{1}{2}({\rm{Re}}f)^{-1 AB}\tilde{\Omega}_{\alpha}k_A^{\alpha}\tilde{\Omega}_{\bar{\beta}}k_B^{\bar{\beta}},\label{comp_HD}
\end{align}
where $\mathcal{A}^2=\mathcal{A}_{\mu}\mathcal{A}^{\mu}$ and $F^0$ is the F-term component of $X^0$.  While $F^0$ is an auxiliary field in the standard supergravity, it becomes a propagating degree of freedom, due to derivative terms. $\tilde{\Omega}_\alpha$ and $\tilde{W}_\alpha$, etc, denote the derivatives with respect to $z^{\alpha}$ or $\bar{z}^{\bar{\alpha}}$, and $\tilde{\Omega}^{\alpha\bar{\beta}}\equiv (\tilde{\Omega}_{\alpha\bar{\beta}})^{-1}$. The derivatives on $z^{\alpha}$ should be understood as the covariant derivatives including gauge connections if $z^{\alpha}$ are charged under some gauge groups, but we omit them for simplicity in the following discussion. $k_A^{\alpha}$ are the Killing vectors defined by the gauge transformations of chiral superfields, $\delta z^{\alpha}=\theta^Ak_A^{\alpha}$, with a transformation parameter $\theta^A$. In the second line of Eq.~$\eqref{comp_HD}$, we find that the Lagrangian contains the non-minimal couplings between matter chiral multiplets and the Ricci scalar, $\tilde{\Omega} (z^{\alpha},\bar{z}^{\bar{\alpha}})R$, and  $R^2$ as desired for the supersymmetric extension of the Higgs-$R^2$ inflation.

In NMSSM, the Higgs sector is composed of
\begin{align}
 z^{\alpha}=\{S,H_{u},H_{d}\}, 
\end{align}
where $S$ is the singlet chiral superfield, and $H_u$ and $H_d$ are $SU(2)_L$ doublet Higgs superfields, given by
\begin{align}
H_{u}=\left(\begin{array}{c}
H_{u}^{+} \\
H_{u}^{0}
\end{array}\right), \quad H_{d}=\left(\begin{array}{c}
H_{d}^{0} \\
H_{d}^{-}
\end{array}\right).    
\end{align}
Then, we choose the frame function and the superpotential, respectively, as~\cite{Einhorn:2009bh,Ferrara:2010yw,Ferrara:2010in},
\begin{align}
&\tilde{\Omega}(z^{\alpha},\bar{z}^{\bar{\beta}})=-3+|S|^2+|H_{u}|^2+|H_{d}|^2+\left(\frac{3}{2}\chi H_u\cdot H_d +{\rm{h.c.}}\right),\label{O_NMSSM}\\
&\tilde{W}(z^{\alpha})=\lambda S H_{u} \cdot H_{d}+\frac{\rho}{3} S^{3},\label{W_NMSSM}
\end{align}
where $|H_{u}|^2=H_{u}^{\dagger} H_{u}$, etc, $H_{u} \cdot H_{d} \equiv-H_{u}^{0} H_{d}^{0}+H_{u}^{+} H_{d}^{-}$, and the frame function is related to the K\"ahler potential  by $\tilde{\Omega}(z^{\alpha},\bar{z}^{\bar{\beta}})=-3\,{\rm exp}\big(-\tilde{\mathcal{K}}(z^{\alpha},\bar{z}^{\bar{\beta}})/3\big)$, and $\chi, \lambda$ and $\rho$ are chosen to be real parameters. 

Therefore, in $R^2$ supergravity, Eq.~$\eqref{comp_HD}$ for the NMSSM describes a supergravity embedding for the system with the non-minimal coupling of the Higgs fields as well as $R^2$ term. The kinetic terms of the NMSSM Lagrangian in $R^2$-supergravity are explicitly given by
\begin{align}
\nonumber \mathcal{L}/\sqrt{-g}=&\left\{\frac{1}{2}-\frac{1}{6}|S|^2-\frac{1}{6}|H_u|^2-\frac{1}{6}|H_d|^2+\left(-\frac{1}{4}\chi H_u\cdot H_d +{\rm{h.c.}}\right)\right\}R\\
&-|\partial_{\mu}S|^2- |\partial_{\mu}H_u|^2- |\partial_{\mu}H_d|^2+\frac{\alpha}{36}R^2+\cdots ,
\end{align}
where the ellipsis denotes the terms containing $\mathcal{A}_{\mu}$, $F^0$ and the scalar potential. Compared to the non-supersymmetric case, we have several additional scalar fields including $S$ and multi-Higgs fields.
We are not going to pursue the above form of the $R^2$-supergravity any longer, but instead we rely on the dual-scalar description of the $R^2$-supergravity in the next subsection.

\subsection{Dual-scalar Lagrangian}\label{sec_dual}

In this section, we derive a dual Lagrangian for Eq.~$\eqref{S_HD}$ by transforming the higher derivative terms such as $R^2$ to dynamical scalar fields including the scalaron. We perform the analysis in a supersymmetric way without imposing a gauge fixing condition on $X^0$ for dilatation. Fixing $X^0$ at a special value corresponds to identifying a frame of the system. In the next section, we define equivalent frames in a unified manner by fixing $X^0$ appropriately. 

Here we derive the master action without specifying $X^0$, following the duality procedure of Ref.~\cite{Cecotti:1987sa} (See also Refs.~\cite{Ferrara:2013wka,Cecotti:2014ipa}). To do so, note that the last term of Eq.~$\eqref{S_HD}$ can be rewritten as 
\begin{align}
[\alpha \bar{\mathcal{R}}\mathcal{R}]_D=[\alpha\bar{C}C]_D+[T(C-\mathcal{R})]_F,  \label{S_HD2}
\end{align}
where $T$ and $C$ are chiral multiplets with weights $(2,2)$ and $(1,1)$, respectively. The EOM of $T$ leads to $C=\mathcal{R}$, and then we can see the equality of Eq.~$\eqref{S_HD2}$. On the other hand, the second term in the right-hand side of Eq.~$\eqref{S_HD2}$ can be transformed as 
\begin{align}
\nonumber [T(C-\mathcal{R})]_F&=[TC-\Sigma(T(X^0)^{-1}\bar{X}^{\bar{0}})]_F\\
&=[TC]_F-[T(X^0)^{-1}\bar{X}^{\bar{0}}+{\rm{c.c.}}]_D,
\end{align}
up to total derivative. Therefore, we obtain the following total action,
\begin{align}
S=[|X^0|^2(\tilde{\Omega}+\alpha\bar{C}C-(T+\bar{T}))]_D+[(X^0)^3(\tilde{W}+TC)]_F+[f_{AB}(z^{\alpha})\bar{\mathcal{W}}^A\mathcal{W}^B]_F,  \label{S_HD3}
\end{align}
where we redefined $T\rightarrow T(X^0)^2$ and $C\rightarrow CX^0$ so that $T$ and $C$ are weightless. 

Comparing Eq.~$\eqref{S_HD3}$ to the standard supergravity, 
\begin{align}
S=[|X^0|^2\Omega (z^I,\bar{z}^{\bar{J}})]_D+[(X^0)^3W(z^{I})]_F+[f_{AB}(z^{I})\bar{\mathcal{W}}^A\mathcal{W}^B]_F,  \label{S_sugra}
\end{align}
we can define new frame function~$\Omega$ and superpotential~$W$,
\begin{align}
\nonumber \Omega(z^{I},\bar{z}^{\bar{J}})&\equiv\tilde{\Omega}(z^{\alpha},\bar{z}^{\bar{\beta}})+|C|^2-(T+\bar{T})\\
&=-3+|S|^2+|H_{u}|^2+|H_{d}|^2+\left(\frac{3}{2}\chi H_u\cdot H_d +{\rm{h.c.}}\right)+|C|^2-(T+\bar{T}),\label{tilde_O}\\
W(z^{I})& \equiv  \tilde{W}(z^{\alpha})+\frac{1}{\sqrt{\alpha}}TC=\lambda S H_{u} \cdot H_{d}+\frac{\rho}{3} S^{3}+\frac{1}{\sqrt{\alpha}}TC,\label{tilde_W}\\
f_{AB}(z^{I})&=f_{AB}(z^{\alpha}),\label{f}
\end{align}
where we redefined $C\rightarrow C/\sqrt{\alpha}$, and the K\"ahler potential is defined by  $\Omega(z^{I},\bar{z}^{\bar{J}})=-3\,{\rm exp}\big(-\mathcal{K}(z^{I},\bar{z}^{\bar{J}})/3\big)$. In this expression, the higher derivative term $\alpha \bar{\mathcal{R}}\mathcal{R}$ disappears, but instead there appear two additional chiral superfields, $T$ and $C$, in the standard supergravity action.  When $\alpha=0$, $C$ appears only in the superpotential, which becomes a Lagrange multiplier forcing $T=0$. 
Then, we recover the original NMSSM inflation model~\cite{Einhorn:2009bh,Ferrara:2010yw,Ferrara:2010in}.

We now derive the bosonic Lagrangian in the general scalar-dual supergravity in detail for the later convenience. 
After imposing the superconformal gauge fixing except for dilatation,\footnote{In particular, $X^0=\bar{X}^{\bar{0}}$ is imposed as $A$-gauge~\cite{Freedman:2012zz}.} and integrating out some auxiliary fields, we obtain
\begin{align}
\nonumber \mathcal{L}/\sqrt{-g}=& -\frac{1}{6}(X^0)^2\Omega R-\Omega (\partial_{\mu}X^0)^2 -X^{0} \partial^{\mu} X^{0}\left(\Omega_{I} \partial_{\mu} z^{I}+\Omega_{\bar{I}} \partial_{\mu} \bar{z}^{\bar{I}}\right)+\left(X^{0}\right)^{2}\Omega\mathcal{A}_{\mu}^2\\
&-(X^0)^2 \Omega_{I\bar{J}} \partial_{\mu} z^{I} \partial^{\mu} \bar{z}^{\bar{J}}-V ,\label{master}
\end{align}
where 
\begin{align}
\mathcal{A}_{\mu}=  -\frac{\mathrm{i}}{2 \Omega}\left(\partial_{\mu} z^{I} \Omega_{I} -\partial_{\mu} \bar{z}^{\bar{I}} \Omega_{\bar{I}} \right). \label{sol_A}
\end{align}
The scalar potential $V=V^F+V^D$ is given by
\begin{align}
\nonumber V^F=&\left(X^{0}\right)^{4} \left(\Omega_{I \bar{J}}-\frac{\Omega_{I} \Omega_{\bar{J}}}{\Omega}\right)^{-1}\left(W_{I}-\frac{3 \Omega_{I}}{\Omega} W\right)\left(\bar{W}_{\bar{J}}-\frac{3\Omega_{\bar{J}}}{\Omega} \bar{W}\right) +\frac{9}{\Omega}\left(X^{0}\right)^{4}|W|^{2} \\
=& \left(X^{0}\right)^{4}e^{\mathcal{K}/3}\left[ \mathcal{K}^{I\bar{J}} \left(W_{I}+\mathcal{K}_IW\right)\left(\bar{W}_{\bar{J}}+\mathcal{K}_{\bar{J}}\bar{W}\right)-3 |W |^2\right],\\
V^D=&\frac{\left(X^{0}\right)^{4}}{2}({\rm{Re}}f)^{-1 AB}\Omega_{\alpha}k_A^{\alpha}\Omega_{\bar{\beta}}k_B^{\bar{\beta}},
\end{align}
where $\mathcal{K}^{I\bar{J}}$ is an inverse of K\"ahler metric.
We note that the above results hold for the general forms of K\"ahler potential (frame function) and the superpotential independently of our choice Eqs.~$\eqref{tilde_O}$-$\eqref{f}$.

For NMSSM in the dual-scalar description of $R^2$-supergravity, henceforth, we use the following notations for the extended Higgs sector,
\begin{align}
&z^I=\left\{S,H_{u},H_{d},C,T\right\},\\
&z^{i}=\{S,H_{u},H_{d},C\},\label{z^i}\\
&z^{\alpha}=\{S,H_{u},H_{d}\}.
\end{align}
Then, from  Eqs.~(\ref{tilde_O}) and (\ref{tilde_W}), the K\"ahler metric and its inverse are simplified due to $\Omega_{i \bar{j}}=\delta_{i \bar{j}}$ and $\Omega_{T \bar{T}}=0$, so they are explicitly given by
\begin{align}
\mathcal{K}_{I\bar{J}}= -\frac{3}{\Omega}\left(\begin{array}{cc}
\delta_{i \bar{j}}-\frac{\Omega_i\Omega_{\bar{j}}}{\Omega} & \frac{\Omega_i}{\Omega} \\
\frac{\Omega_{\bar{j}}}{\Omega}  & - \frac{1}{\Omega}
\end{array}\right), \ \   \mathcal{K}^{I\bar{J}}=-\frac{\Omega}{3} \left(\begin{array}{cc}
 \delta^{i\bar{j} }& \delta^{i\bar{k}}\Omega_{\bar{k}}\\
\delta^{\bar{j}\ell}\Omega_{\ell}& -\Omega+\delta^{k\bar{\ell}} \Omega_{k} \Omega_{\bar{\ell}}
\end{array}\right).\label{metric}
\end{align}
Then, we find that the $F$-term scalar potential can be rewritten as
\begin{align}
V^F=\left(X^{0}\right)^{4}\left[\delta^{i \bar{j}} W_{i} \bar{W}_{\bar{j}}+\left(W_i\delta^{i \bar{j}} \Omega_{\bar{j}}  \bar{W}_{\bar{T}}-3W_{T} \bar{W}+{\rm{c.c}}\right)-\left(\Omega-\delta^{i \bar{j}}{\Omega}_{i} \Omega_{\bar{j}}\right)|W_{T}|^{2}\right].\label{SP_1}
\end{align}
Therefore, in the absence of the chiral superfields $T$ and $C$, the above $F$-term potential takes the same form as in the NMSSM with global SUSY when we take $X^0=1$ (Jordan frame).
Otherwise, the scalar potential deviates from the one in the NMSSM and there appears an important contribution from $T$ for inflation as will be shown in the later sections.


\section{Dual-scalar supergravity for NMSSM} \label{S2}

\begin{figure}[t]

  \begin{center}
   \includegraphics[width=120mm]{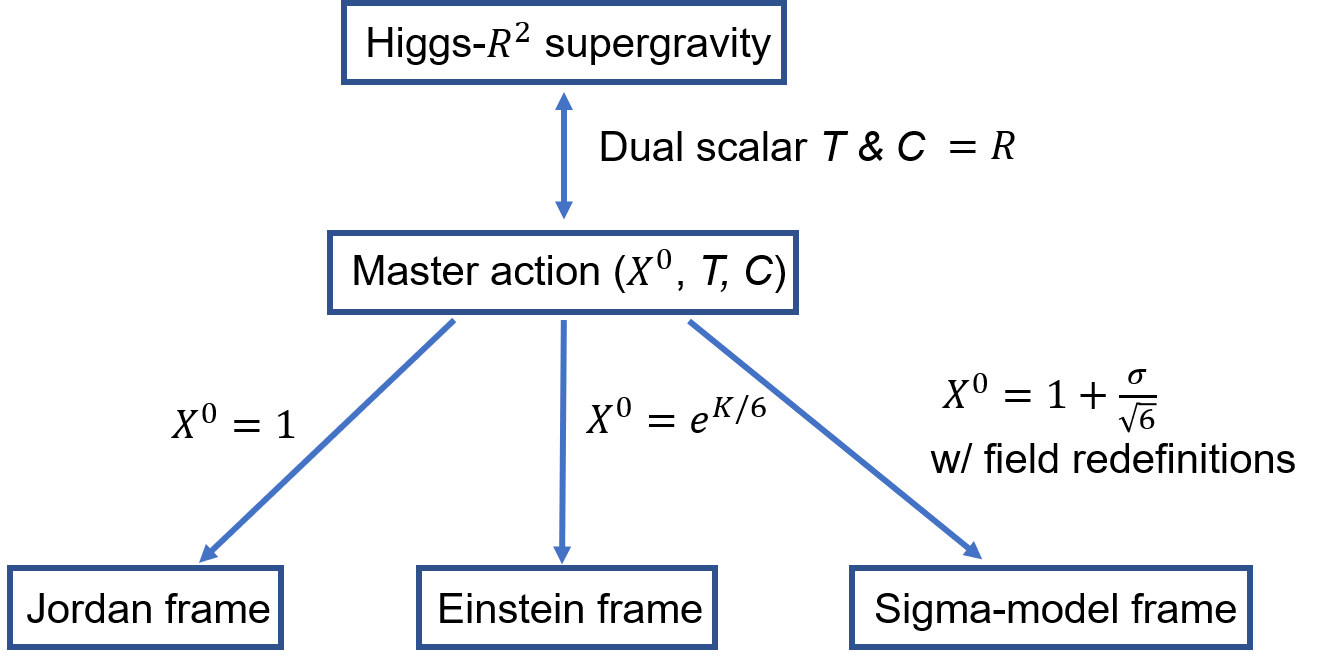}
  \end{center}
\caption{Relation between different frames}
  \label{fig:structure}
 
\end{figure}

In this section, we introduce equivalent frames by choosing certain gauge conditions for superconformal symmetry and discuss the universal properties of Higgs-$R^2$ supergravity in more detail. After the dual transformation of the $R^2$ term in superspace, we introduce three different kinds of frames (Jordan frame, Einstein frame, and linear sigma frame). We put the equivalent frames in the unified fashion and distinguish our work from the previous ones by theoretical constraints such as perturbative bounds on the parameters, unitarity, and inflation dynamics, etc.

\subsection{Jordan frames}\label{CJF}

We first take $X^0=1$ in Eq.~$\eqref{master}$, for which the bosonic Lagrangian becomes
\begin{align}
\mathcal{L}_J/\sqrt{-g}=& -\frac{1}{6}\Omega R- \Omega_{I\bar{J}} \partial_{\mu} z^{I} \partial^{\mu} \bar{z}^{\bar{J}}-V_J,
\end{align}
where we omit the term with $\mathcal{A}_{\mu}$.\footnote{In the most of situations, this term vanishes at the inflation background.} Again, this expression holds for a general system with frame function and superpotential independently of our setup. Then, if the frame function takes a form,
\begin{align}
\Omega=-3e^{-\frac{\mathcal{K}}{3}}= -3 +\delta_{I\bar{J}}z^I\bar{z}^{\bar{J}}+J(z)+\bar{J}(\bar{z}), \label{CJ}   
\end{align}
where $J$ is an arbitrary holomorphic function, the scalar fields have canonical kinetic terms~\cite{Ferrara:2010yw,Ferrara:2010in}}, and we call this frame as a canonical Jordan frame.\footnote{Moreover, when $J=0$ and the superpotential contains only cubic terms, the scalar potential becomes the same as that of the global supersymmetric theory. This kind of model is named as ``canonical superconformal supergravity"~\cite{Ferrara:2010in}.}  However, we note that there is a slight difference between Eq.~$\eqref{CJ}$ and the frame function in Eq.~$\eqref{tilde_O}$: the dual scalar field $T$ appears as $T+\bar{T}$ in $\Omega$, so it does not have a kinetic term in the Jordan frame. Although the $T$-dependent frame function in Eq.~$\eqref{CJ}$ are specific to the $R^2$ supergravity \footnote{We remind ourselves that $T$ is introduced as a Lagrange multiplier in the dual process.}, we keep the terminology, ``Jordan frame", in this paper, in order to refer to the matter part of the frame function.

Taking Eqs.~$\eqref{tilde_O}$ and $\eqref{tilde_W}$ in Jordan frame, we obtain the following bosonic part of the Lagrangian:
\begin{align}
\nonumber \mathcal{L}_J/\sqrt{-g}=&\left\{\frac{1}{2}-\frac{1}{6}|S|^2-\frac{1}{6}|H_u|^2-\frac{1}{6}|H_d|^2-\frac{1}{6}|C|^2+\left(-\frac{1}{4}\chi H_u\cdot H_d +{\rm{h.c.}}\right)+\frac{1}{3}{\rm{Re}}T\right\}R\\
&- |\partial_{\mu}S|^2- |\partial_{\mu}H_u|^2- |\partial_{\mu}H_d|^2- |\partial_{\mu}C|^2+\Omega\mathcal{A}_{\mu}^2-V_J.\label{dual2}
\end{align}
Here, the scalar potential $V_J=V_J^F+V_J^D$ is given by
\begin{align}
\nonumber V_{J}^F=&\left|\lambda H_u\cdot H_d+\rho S^{2}\right|^{2}+\lambda^2\left| S \right|^{2}(|H_u|^2+|H_d|^2)+\frac{1}{\alpha}|T|^2 \nonumber \\
&+\frac{3}{2}\frac{\chi\lambda}{\sqrt{\alpha}} (S\bar{C}+\bar{S}C)(|H_u|^2+|H_d|^2) \nonumber \\
&+\frac{1}{\alpha}|C|^2\left\{3+\frac{3}{2}\chi( H_u\cdot H_d+{\rm{c.c.}})+\frac{9}{4}\chi^2(|H_u|^2+|H_d|^2)-2{\rm{Re}}T\right\},\\\label{SP_1explicit}
V_{J}^{D}=&\frac{g^{\prime 2}}{8}\left(\left|H_{u}\right|^{2}-\left|H_{d}\right|^{2}\right)^{2}+\frac{g^{2}}{8}\left(\left(H_{u}\right)^{\dagger} \vec{\tau} H_{u}+\left(H_{d}\right)^{\dagger} \vec{\tau} H_{d}\right)^{2}.  
\end{align}
where ${\tau_i} (i=1,2,3)$ are Pauli matrices. 
For the $D$-term scalar potential, we took $f_{AB}=\delta_{AB}$ and kept the part for $U(1)_Y$ and $SU(2)_L$ groups with the corresponding  gauge couplings, $g'$ and $g$, respectively.
Therefore, we have obtained the generalized NMSSM inflation model in Jordan frame with additional two complex fields ($C$ and $T$). We note that there is a non-minimal coupling for ${\rm{Re}}T$ with $R$, but not for ${\rm{Im}}T$.

We now consider the limit of $\alpha\rightarrow 0$, for which the Starobinsky corrections disappear. After redefining  $C\rightarrow\sqrt{\alpha}C$ and $T\rightarrow\sqrt{\alpha}T$, and taking the limit $\alpha\rightarrow 0$, $C$ and $T$ do not appear in the frame function, so they become auxiliary fields. Then, after integrating out them by using the equations of motion, we obtain
\begin{align}
T=0, \ \ C=\frac{\frac{3}{2} \chi \lambda  S\left(|H_u|^{2}+|H_d|^{2}\right)}{-3+\left(-\frac{3}{2}\chi H_{u}\cdot H_{d}+{\rm{c.c.}}\right)-\frac{9}{4} \chi^{2}\left(|H_u|^{2}+|H_d|^{2}\right)}.    
\end{align}
Then, plugging the above relations back to Eq.~$\eqref{SP_1explicit}$, we get 
\begin{align}
\nonumber V_{J}^F|_{\alpha\rightarrow 0}=&\left|\lambda H_u\cdot H_d+\rho S^{2}\right|^{2}\\
&+\lambda^{2}|S|^{2}\left(|H_u|^{2}+|H_d|^{2}\right) \frac{-3+\left(-\frac{3}{2} \chi H_{u}\cdot H_{d}+{\rm{c.c.}}\right)}{-3+\left(-\frac{3}{2}\chi H_{u}\cdot H_{d}+{\rm{c.c.}}\right)-\frac{9}{4} \chi^{2}\left(|H_u|^{2}+|H_d|^{2}\right)},
\end{align}
which reproduces the result of Refs.~\cite{Ferrara:2010yw,Ferrara:2010in}. On the other hand, if $\alpha$ is sizable, the additional scalar fields $C, T$ become dynamical, so we need to take them into account for inflationary dynamics. 
For a conformal coupling for the Higgs fields, i.e. $\chi=0$, the NMSSM sector is decoupled from the scalaron, so we recover the pure Starobinsky inflation in supergravity~\cite{Cecotti:1987sa}. Otherwise, our model interpolates between Higgs and Starobinsky inflation models in supergravity.  

We remark that setting $X^0=\sqrt{-3/\Omega}=e^{\mathcal{K}/6}$ in Eq.~$\eqref{master}$ leads to the dual-scalar Lagrangian for  NMSSM in Einstein frame, as follows,
\begin{align}
\mathcal{L}_E/\sqrt{-g}=\frac{1}{2}R-\mathcal{K}_{I \bar{J}} \partial_{\mu} z^{I} \partial^{\mu} \bar{z}^{\bar{J}}-V_E, \label{E-frame}  
\end{align}
where the scalar potential~$V_E$ is related to $V_{J}$ as
\begin{align}
\nonumber V_E=&\frac{9}{\Omega^2}V_{J}.\label{V_E}
\end{align}
We note that the kinetic term for $T$ is contained in Eq.~$\eqref{E-frame}$, unlike the canonical Jordan frame in Eq.~(\ref{dual2}).


\subsection{Sigma-model frames}\label{LSF}

Next we introduce ``linear sigma frame", where it is easy to see the recovery of unitarity problem.  
To do so, we redefine the matter multiplets as
\begin{align}
\hat{z}^i\equiv X^0 z^i, \ \ \hat{T}\equiv  (X^0)^2 T, 
\end{align}
with $z^{i}=\{S,H_{u},H_{d},C\}$. Then, the frame function and the superpotential can be rewritten as
\begin{align}
|X^0|^2\Omega (z^I,\bar{z}^{\bar{J}})
&= -3 |X^0|^2+|\hat{S}|^2+|\hat{H}_{u}|^2+|\hat{H}_{d}|^2+|\hat{C}|^2 \nonumber \\
&\quad+\frac{3\chi}{2}\left( \frac{\hat{H}_u\cdot \hat{H}_d\bar{X}^{\bar{0}}}{X^0} +{\rm{h.c.}}\right)-\left( \frac{\hat{T}\bar{X}^{\bar{0}}}{X^0} +{\rm{h.c.}}\right),\\
(X^0)^3W(z^I)&=\lambda \hat{S} \hat{H}_{u} \cdot \hat{H}_{d}+\frac{\rho}{3} \hat{S}^{3}+\frac{1}{\sqrt{\alpha}}\hat{T}\hat{C}.
\end{align}
After imposing gauge fixing conditions except for the dilatation and integrating out auxiliary fields, Equation~$\eqref{S_HD3}$ produces the following bosonic terms,
\begin{align}
 \mathcal{L}_{LS}/\sqrt{-g}&=\bigg\{\frac{(X^0)^2}{2}-\frac{1}{6}|\hat{S}|^2-\frac{1}{6}|\hat{H}_u|^2-\frac{1}{6}|\hat{H}_d|^2-\frac{1}{6}|\hat{C}|^2 \nonumber \\
&\qquad+\left(-\frac{1}{4}\chi \hat{H}_u\cdot \hat{H}_d +{\rm{h.c.}}\right)+\frac{1}{3}{\rm{Re}}\hat{T}\bigg\}R \nonumber \\
&\quad+\left(\left(\partial \log X^0\right)^{2}+\Box \log X^0\right)\left(-3(X^0)^2+\left(\frac{3}{2}\chi \hat{H}_u\cdot \hat{H}_d +{\rm{h.c.}}\right)-2{\rm{Re}}\hat{T}\right) \nonumber  \\
&\quad- |\partial_{\mu}\hat{S}|^2- |\partial_{\mu}\hat{H}_u|^2- |\partial_{\mu}\hat{H}_d|^2- |\partial_{\mu}\hat{C}|^2+\Omega\mathcal{A}_{\mu}^2-V_{LS},
\end{align}
where
\begin{align}
\nonumber V_{LS}^F=&|\lambda \hat{H}_u\cdot \hat{H}_d+\rho \hat{S}^{2}|^{2}+\lambda^2| \hat{S} |^{2}(|\hat{H}_u|^2+|\hat{H}_d|^2)+\frac{1}{\alpha}|\hat{T}|^2 \nonumber \\
&+\frac{3}{2}\frac{\chi\lambda}{\sqrt{\alpha}} (\hat{S}\bar{\hat{C}}+\bar{\hat{S}}\hat{C})(|\hat{H}_u|^2+|\hat{H}_d|^2) \nonumber \\
&-\frac{1}{\alpha}|\hat{C}|^2\left\{-3(X^0)^2-\frac{3}{2}\chi( \hat{H}_u\cdot \hat{H}_d+{\rm{c.c.}})-\frac{9}{4}\chi^2(|\hat{H}_u|^2+|\hat{H}_d|^2)+2{\rm{Re}}\hat{T}\right\},\label{V_LS} \\
V_{LS}^{D}=&\frac{g^{\prime 2}}{8}\left(|\hat{H}_{u}|^{2}-|\hat{H}_{d}|^{2}\right)^{2}+\frac{g^{2}}{8}\left((\hat{H}_{u})^{\dagger} \vec{\tau} \hat{H}_{u}+(\hat{H}_{d})^{\dagger} \vec{\tau} \hat{H}_{d}\right)^{2}.
\end{align}
Here, we note that $\mathcal{A}_{\mu}$ in the sigma-model frame is shown explicitly in the new basis, as follows,
\begin{align}
\mathcal{A}_{\mu}= -\frac{i}{2\Omega}(X^{0})^2\left(\Omega_{\hat{i}}\partial_{\mu}((X^{0})^{-1}\hat{z}^{\hat{i}})+X^0\Omega_{\hat{T}}\partial_{\mu}((X^{0})^{-2}\hat{T})-{\rm{c.c.}}\right). \label{Amu}
\end{align}
Now, we fix the dilatation gauge by $X^0=1+\frac{1}{\sqrt{6}} \sigma$, where $\sigma$ is a function of $\hat{z}^I$ and $\bar{\hat{z}}^{\bar{J}}$. In particular, we choose $\sigma$ so that it satisfies the following constraint~\cite{Lee:2021dgi},
\begin{align}
\nonumber &\frac{(X^0)^2}{2}-\frac{1}{6}|\hat{S}|^2-\frac{1}{6}|\hat{H}_u|^2-\frac{1}{6}|\hat{H}_d|^2-\frac{1}{6}|\hat{C}|^2+\left(-\frac{1}{4}\chi \hat{H}_u\cdot \hat{H}_d +{\rm{h.c.}}\right)+\frac{1}{3}{\rm{Re}}\hat{T}\\ &=\frac{1}{2}-\frac{1}{6}|\hat{S}|^2-\frac{1}{6}|\hat{H}_u|^2-\frac{1}{6}|\hat{H}_d|^2-\frac{1}{6}|\hat{C}|^2-\frac{1}{12}\sigma^2, \label{frameredef}
\end{align}
or equivalently
\begin{align}
\left(1+\frac{1}{\sqrt{6}} \sigma\right)^2 +\left(-\frac{1}{2}\chi \hat{H}_u\cdot \hat{H}_d +{\rm{h.c.}}\right) +\frac{2}{3}{\rm{Re}}\hat{T} =1-\frac{1}{6}\sigma^2.\label{def_sigma}
\end{align}
By this equation, $X^0$ (or $\sigma$) can be expressed in terms of only physical fields $\hat{z}^I$. Instead of doing so, we use the equation to eliminate ${\rm{Re}}\hat{T}$ regarding $\sigma$ as a new dynamical field. Then, we obtain
\begin{align}
\nonumber \mathcal{L}_{LS}/\sqrt{-g}=&\frac{1}{2}\left(1-\frac{1}{3}|\hat{S}|^2-\frac{1}{3}|\hat{H}_u|^2-\frac{1}{3}|\hat{H}_d|^2-\frac{1}{3}|\hat{C}|^2-\frac{1}{6}\sigma ^2\right)R\\
&- |\partial_{\mu}\hat{S}|^2- |\partial_{\mu}\hat{H}_u|^2- |\partial_{\mu}\hat{H}_d|^2- |\partial_{\mu}\hat{C}|^2-\frac{1}{2}(\partial_{\mu}\sigma)^2 +\Omega\mathcal{A}_{\mu}^2-V_{LS},\label{LS_frame}
\end{align}
where the scalar potential~$\eqref{V_LS}$ becomes
\begin{align}
\nonumber V_{LS}^F=&|\lambda \hat{H}_u\cdot \hat{H}_d+\rho \hat{S}^{2}|^{2}+\lambda^2| \hat{S} |^{2}(|\hat{H}_u|^2+|\hat{H}_d|^2) \nonumber \\
&+ \frac{1}{4\alpha}\left(\sigma^2+\sqrt{6}\sigma-\left(\frac{3}{2}\chi \hat{H}_u\cdot \hat{H}_d +{\rm{h.c.}}\right)\right)^2 \nonumber \\
\nonumber &+ \frac{1}{\alpha}({\rm{Im}}\hat{T})^2+\frac{3}{2}\frac{\chi\lambda}{\sqrt{\alpha}} (\hat{S}\bar{\hat{C}}+\bar{\hat{S}}\hat{C})(|\hat{H}_u|^2+|\hat{H}_d|^2) \nonumber \\
&+\frac{1}{\alpha}|\hat{C}|^2\left\{3+2\sqrt{6}\sigma+\frac{3}{2}\sigma^2+\frac{9}{4}\chi^2(|\hat{H}_u|^2+|\hat{H}_d|^2)\right\}.\label{SP_2explicit}
\end{align}
Taking into account the extra kinetic terms coming from $\mathcal{A}^2_{\mu}$ with Eq.~(\ref{Amu}) in Eq.~(\ref{LS_frame}), we find that all the scalar fields including ${\rm Im}\,{\hat T}$ turn out to be dynamical and there is no unitarity violation up to the Planck scale after the field redefinitions, as will be discussed shortly below.   Indeed, as shown in Ref.~\cite{Lee:2021dgi}, it is obvious that the scalaron $\sigma$ plays a role of a sigma field in the linear sigma model, which pushes up the unitarity bound to the Planck scale.
Moreover, the local conformal invariance is respected in Eq.~$\eqref{LS_frame}$, except for the Planck mass and the scalar potential.

Before closing the section, we remark on the angular part of the complex scalar fields and the unitarity problem in more detail. 
As commented above, in the sigma-model frame, the kinetic terms for the angular part of the complex scalar fields come from $\Omega A^2_\mu$ in Eq.~(\ref{LS_frame}) with Eq.~(\ref{Amu}). We enumerate them in the following,
\bea
\Omega {\cal A}^2_\mu &=&-\frac{1}{4\Omega}\, (X^0)^4 \bigg[(X^0)^{-1} (\Omega_{\hat i} \partial_\mu {\hat z}^{\hat i}+\Omega_{\hat T}\partial_\mu {\hat T})+(\partial_\mu (X^0)^{-1}) (\Omega_{\hat i} {\hat z}^{\hat i} +2 \Omega_{\hat T}{\hat T})-{\rm c.c.}\bigg]^2 \nonumber \\
&=&-\frac{1}{4\Omega}\, \bigg[ (X^0)^{-1}\Big(({\bar {\hat C}}\partial_\mu {\hat C} +{\bar {\hat S}}\partial_\mu {\hat S}+{\bar {\hat H}}_u \partial_\mu {\hat H}_u+{\bar {\hat H}}_d \partial_\mu {\hat H}_d-{\rm c.c}) -2i \partial_\mu b \Big) \nonumber \\
&&\quad-4ib\,\partial_\mu (X^0)^{-1} \bigg]^2.
\eea
Here, we have redefined ${\hat T}-{\bar {\hat T}}$ in terms of a real scalar field $b$ as
\bea
{\hat T}-{\bar {\hat T}}-\frac{3}{2}\chi ({\hat H}_u \cdot {\hat H}_d -{\bar {\hat H}}_u\cdot {\bar {\hat H}_d}) =2ib,
\eea 
and we note that the frame function can be written with the constraint in Eq.~(\ref{frameredef})  as
\bea
\Omega=-6(X^0)^{-2} \bigg(\frac{1}{2}-\frac{1}{6}|\hat{S}|^2-\frac{1}{6}|\hat{H}_u|^2-\frac{1}{6}|\hat{H}_d|^2-\frac{1}{6}|\hat{C}|^2-\frac{1}{12}\sigma^2\bigg).
\eea
Therefore, the mass term for ${\rm Im {\hat T}}$ in the scalar potential (\ref{SP_2explicit}) becomes
\bea
V_{LS}^F\supset \frac{1}{\alpha}({\rm{Im}}\hat{T})^2=  \frac{1}{\alpha}\bigg(b-\frac{3}{4}i \chi  ({\hat H}_u\cdot {\hat H}_d -{\bar {\hat H}}_u \cdot {\bar {\hat H}_d}) \bigg)^2.
\eea
As a result, in the resulting Lagrangian in the sigma-model frame, the non-minimal coupling $\chi$ for the Higgs fields appearing in $\Omega {\cal A}^2_\mu$ has been moved to the scalar potential so there is no large coupling in the kinetic terms. This is similar to the previous observation that the constraint  for ${\hat T}+{\bar {\hat T}}$ in Eq.~(\ref{frameredef}) is imposed  in the sigma-model frame for eliminating the non-minimal coupling appearing in the frame function $\Omega$.
In this case, we only have to impose the perturbativity  bounds on the couplings between $b$ and the Higgs fields from the redefined mass term, as follows,
\bea
\frac{\chi}{\alpha}\lesssim 1, \qquad  \frac{\chi^2}{\alpha}\lesssim 1.
\eea
 
 Moreover, as will be shown in the next section, the angular parts of the complex scalar fields are decoupled during inflation, so they are not relevant for our inflation discussion.


\section{Higgs-Sigma inflation}\label{S3}
We now apply our results on dual-scalar description of Higgs-$R^2$ supergravity and apply them for Higgs-sigma inflation. We first derive the effective action for a slow-roll inflation in the Higgs-sigma system and then show the necessary conditions for the stability of the inflationary trajectory.

\subsection{Effective action for inflation}
First we consider the effective action for inflation in Jordan frame supergravity, written in terms of the original variables used in Sec.~\ref{CJF}.
To that, keeping only the scalaron ${\rm{Re}}T$ and  the neutral Higgs field $h$ from $H_u^{0}\rightarrow \frac{1}{2}h$ and $H_d^{0}\rightarrow \frac{1}{2}h$, and setting all the other fields to zero, we obtain the following Lagrangian in the Einstein frame:  
\begin{align}
\nonumber \mathcal{L}/\sqrt{-g}=&\frac{1}{2}R-\frac{1}{2}\frac{\big(1+\xi (1+6\xi)h^2+\frac{2}{3}{\rm{Re}}T\big)}{(1+\xi h^2+\frac{2}{3}{\rm{Re}}T)^2}(\partial_{\mu}h)^2 -\frac{1}{3}\frac{1}{(1+\xi h^2+\frac{2}{3}{\rm{Re}}T)^2}(\partial_{\mu}{\rm{Re}}T)^2\\
&-\frac{2\xi h}{(1+\xi h^2+\frac{2}{3}{\rm{Re}}T)^2}\partial_{\mu}h\partial^{\mu}{\rm{Re}}T-V(h,{\rm{Re}}T),
\end{align}
where the effective non-minimal coupling for the Higgs field is given by 
\begin{align}
\xi \equiv -\frac{1}{6}+\frac{\chi}{4},
\end{align}
and the Einstein frame scalar potential is
\begin{align}
V(h,{\rm{Re}}T)=\frac{1}{(1+\xi h^2+\frac{2}{3}{\rm{Re}}T)^2}\left(\frac{1}{16} \lambda^{2} h^{4}+\frac{1}{\alpha}({\rm{Re}}T)^{2}\right).    
\end{align}
The obtained Lagrangian is equivalent to the one in the non-supersymmetric Higgs-$R^2$ inflation in Ref.~\cite{Ema:2017rqn,Ema:2020zvg,Lee:2021dgi}. 

We can take the alternative basis for fields in sigma model frame, written in terms of the rescaled fields, ${\hat z}^i=X^0 z^i$ and ${\hat T}=(X^0)^2 T$,  introduced in Sec.~\ref{LSF}. Then, using $\hat{h}=X^0h=\left(1+\frac{1}{\sqrt{6}} \sigma\right)h$ and redefining ${\rm Re} {\hat T}$ in terms of the $\sigma$ field satisfying Eq.~$\eqref{def_sigma}$, we can rewrite the Lagrangian in Einstein frame as \begin{align}
\nonumber \mathcal{L}/\sqrt{-g}=&\frac{1}{2}R-\frac{1}{2}\frac{1}{\left(1-\frac{1}{6}\hat{h}^2-\frac{1}{6}\sigma^2\right)^2}\biggl[\left(1-\frac{\sigma^2}{6}\right)(\partial_{\mu}\hat{h})^2+\left(1-\frac{\hat{h}^2}{6}\right)(\partial_{\mu}\sigma)^2\\
&+\frac{1}{3}\hat{h}\sigma \partial_{\mu}\hat{h}\partial^{\mu}\sigma\biggr]-V(\hat{h},\sigma),\label{L_inf_hat}
\end{align}
where 
\begin{align}
V(\hat{h},\sigma)=\frac{1}{\left(1-\frac{1}{6}\hat{h}^2-\frac{1}{6}\sigma^2\right)^2} \biggl[\frac{\lambda^2}{16}\hat{h}^4+\frac{1}{4\alpha}\left(3\left(\xi+\frac{1}{6}\right)\hat{h}^2+\sqrt{6}\sigma+\sigma^2\right)^2\biggr].\label{V_h_sigma}
\end{align}
The inflationary dynamics in this picture is studied in Ref.~\cite{Lee:2021dgi}, and we adopt the above basis for inflation in the following discussion. 

As a result, we find that a non-minimal coupling for the Higgs fields contributes to the extra quartic coupling for the Higgs fields and the mixing quartic coupling between the Higgs and sigma fields, so we only have to impose  the perturbativity bounds on them \footnote{We can compare with the parameters in Ref.~\cite{Lee:2021dgi} by $\alpha \rightarrow 1/\kappa_1$ and $\lambda\rightarrow 2\sqrt{\lambda}$. We note that in NMSSM, the Higgs quartic coupling in the vacuum not only contains $\lambda$, but the electroweak gauge couplings, $g$ and $g'$.}, as follows,
\begin{align}
\frac{\lambda^2}{4}+ \frac{9}{\alpha}\left(\xi+\frac{1}{6}\right)^2 \leq 1, \ \   0<\frac{1}{\alpha}
\leq 1, \ \ \frac{6}{\alpha}\left(\xi+\frac{1}{6}\right)\leq 1.\label{perturbativity}
\end{align}
Therefore, even for a large non-minimal coupling for the Higgs fields in the Jordan frame supergravity, the unitarity can be ensured up to the Planck scale due to the sigma field couplings \cite{Ema:2020zvg,Lee:2021dgi}.

In order to obtain the effective inflaton potential for $\sigma$, we integrate out $\hat{h}$.
Thus, ignoring the kinetic terms for the Higgs fields and using the equation of motion, we obtain a solution  for $\hat{h}$ \cite{Lee:2021dgi} as
\begin{align}
\hat{h}^2= \frac{\frac{1}{\alpha} \sigma(\sigma+\sqrt{6})\left(\sigma-3\left(\xi+\frac{1}{6}\right)(\sigma-\sqrt{6})\right)}{\frac{\lambda^2}{4}(\sigma-\sqrt{6})- \frac{3}{\alpha}\left(\xi+\frac{1}{6}\right)\left(\sigma-3\left(\xi+\frac{1}{6}\right)(\sigma-\sqrt{6})\right)}. \label{hath}  
\end{align}
Then, plugging the above solution back to the potential in Eq.~(\ref{V_h_sigma}), we obtain the effective scalar potential  for inflaton $\sigma$,
\begin{align}
V_{\rm{eff}}(\sigma)= \frac{9\lambda^2}{4\alpha}\sigma^2\left[\frac{\lambda^2}{4}(\sigma-\sqrt{6})^2+\frac{1}{\alpha}\left(\sigma-3\left(\xi+\frac{1}{6}\right)(\sigma-\sqrt{6})\right)^2\right]^{-1}. 
\end{align}
In terms of the approximate canonical field for inflaton, $\phi$, related to the sigma field by
\begin{align}
\sigma\simeq -\sqrt{6} \tanh \left(\frac{\phi}{\sqrt{6}}\right),   \label{def_phi}
\end{align}
we finally reach the effective potential for inflaton, 
\begin{align}
 V_{\rm{eff}}(\phi)= \frac{9}{4\alpha}  \left(1-e^{- \frac{2}{\sqrt{6}}\phi}\right)^{2}\left[1+\frac{1}{\lambda^{2} \alpha}\left(6 \xi+e^{- \frac{2}{\sqrt{6}}\phi}\right)^{2}\right]^{-1}.\label{V_inf}
\end{align}
As a consequence, we can recover the pure $R^2$ inflation for $\frac{\xi^2}{\alpha}\ll \lambda^2$ or the Higgs inflation for $\frac{\xi^2}{\alpha}\gg \lambda^2$, in the following way,
\begin{align}
V_{\operatorname{eff}}(\phi) \approx \begin{cases}\frac{9 }{4\alpha}\left(1-e^{-2 \phi / \sqrt{6}}\right)^{2}, & \frac{\xi^2}{\alpha}\ll \lambda^2  \\ \frac{\lambda^2}{16 \xi^{2}}\left(1-e^{-2 \phi / \sqrt{6}}\right)^{2}. & \frac{\xi^2}{\alpha}\gg \lambda^2\end{cases}    
\end{align}
We note that we used the approximations for the fields during inflation, for which Eqs.~$\eqref{def_phi}$ and $\eqref{hath}$ become
\begin{align}
&\sigma \simeq -\sqrt{6}\left(1-2e^{-\frac{2}{\sqrt{6}}\phi}\right),\label{sigma_app}\\
&h^2\simeq \frac{144\frac{\xi}{\alpha}}{\lambda^2+6\frac{\xi}{\alpha}(6\xi+1)}e^{-\frac{2}{\sqrt{6}}\phi}, \label{h_app}   
\end{align}
These approximations are taken for $\phi\gg 1$.
Thus, it is justified that the contribution of the Higgs field  to the kinetic term for $\sigma$ is negliglble for both $R^2$-like inflation and Higgs-like inflation.

During inflation (for $\phi\gg 1$), the inflaton potential in Eq.~(\ref{V_inf}) becomes very flat, so the slow-roll inflation with a single field is realized.
Then, the inflationary observables, $n_s$ and $r$, are given in terms of the number of efoldings $N$ and the parameters in the inflaton potential~\cite{Lee:2021dgi}, as follows,
\bea
n_s &=& 1-\frac{2}{N} -\frac{9}{2N^2} + \frac{3}{\alpha N^2}\, \frac{(-\lambda^2+12\lambda^2\xi+72\xi^2(1+6\xi)/\alpha)}{(\lambda^2+6\xi(1+6\xi)/\alpha)^2}
\eea
and 
\bea
r= 16\epsilon_* = \frac{12}{N^2}. 
\eea
The results agree with the Planck data within $1\sigma$ \cite{Planck:2018jri}. 
Moreover, from the inflation scale,
\begin{align}
V_0\simeq \frac{\frac{9\lambda^2}{4\alpha}}{\lambda^2+36\frac{\xi^2}{\alpha}} =3H^2, \label{Inf_scale}
\end{align}
we also impose the CMB normalization of the power spectrum to get the following constraint on the model parameters,
\begin{align}
\frac{\lambda^2+\frac{36\xi^2}{\alpha}}{\lambda^2/\alpha} =2.25\times 10^{10}.  \label{CMB} 
\end{align}

\begin{figure}[t]
 \begin{minipage}{0.5\hsize}
  \begin{center}
   \includegraphics[width=70mm]{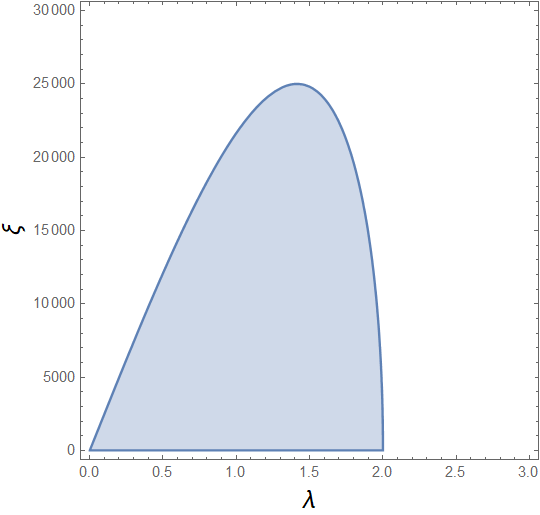}
  \end{center}
 \end{minipage}
 \begin{minipage}{0.5\hsize}
  \begin{center}
   \includegraphics[width=70mm]{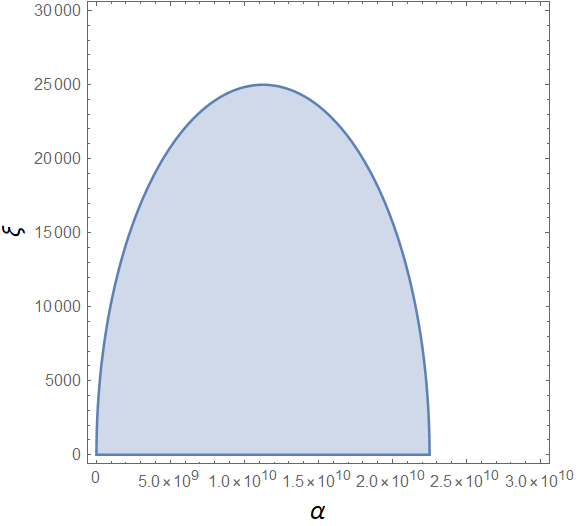}
  \end{center}
 \end{minipage}

  \caption{The region allowed by perturbativity on $\lambda,\xi, $ and $\alpha$.  We used the CMB normalization to express $\alpha (\lambda)$ in terms of the other parameters on the left (right) figure. }
  \label{fig:perturbativity}
\end{figure}

In Fig.~\ref{fig:perturbativity}, eliminating one of the parameters among $\lambda$, $\alpha$, and $\xi$, with Eq.~(\ref{CMB}) and taking into account the perturbativity conditions~$\eqref{perturbativity}$, we show the allowed parameter space for the remaining parameters. From the results, we find the representative values for the parameters, $(\lambda,\xi,\alpha)\sim(0.5,10^4,10^{10})$ and $(\lambda,\xi,\alpha)\sim(4\times 10^{-5},1,10^{2})$, which correspond to the $R^2$-like and Higgs-like inflations, respectively. 

Comments on reheating in our scenario is in order. 
It is important to understand the reheating dynamics for determining the reheating temperature and for the production of dark matter, etc.  There have been recent discussions on the reheating in the context of pure Starobinsky inflation with supergravity~\cite{Terada:2014uia} or without supergravity \cite{Cheong:2020rao} or in the non-supersymmetric Higgs-$R^2$ inflation~\cite{He:2018mgb,He:2020qcb}.
So, it would be interesting to compare our model with those in the literature and discuss the effects of extra scalars and supersymmetric particles during reheating.

\subsection{Decoupling of heavy scalars}
In the previous subsection, we assumed that all the fields other than the radial components of neutral Higgs fields and the scalaron are stabilized at the origin with sufficiently large masses. In this subsection, in order to ensure the stability of the inflaton, we check the decoupling of non-inflaton fields explicitly by analyzing the full scalar potential in the Einstein frame. We use $\hat{z}^i$ and $\sigma$ as fundamental variables, and omit the hat in this subsection.

We first parametrize the MSSM Higgs fields in the following,
\begin{align}
H_u^0=\frac{1}{\sqrt{2}}h\,{\rm{cos}}\beta \, e^{i\delta_1},\ \ H_d^0=\frac{1}{\sqrt{2}}h\, {\rm{sin}}\beta\,  e^{i\delta_2} 
\end{align}
where $h,\beta,\delta_{1,2}$ are all real. Then, the scalar potential in Einstein frame can be rewritten as 
\begin{align}
V_E=\frac{V_{LS}}{\left(1-\frac{1}{3}|S|^2-\frac{1}{6} h^{2}-\frac{1}{3}|H_u^+|^2-\frac{1}{3}|H_d^-|^2-\frac{1}{3}|C|^2-\frac{1}{6} {\sigma}^{2}\right)^2} \label{full_V}   
\end{align}
where the scalar potential in the sigma model frame is 
\begin{align}
\nonumber V_{LS}=&\left|-\frac{1}{4} \lambda h^{2} \sin 2 \beta e^{i \gamma}+\lambda H_{u}^{+} H_{d}^{-}+\rho S^{2}\right|^{2}+\lambda^{2}|S|^{2}\left(\frac{1}{2} h^{2}+\left|H_{u}^{+}\right|^{2}+\left|H_{d}^{-}\right|^{2}\right)\\
\nonumber&+\frac{1}{4\alpha}\left(\sigma^2+\sqrt{6}\sigma+\frac{3}{4}\chi h^2 \sin 2\beta \cos \gamma -\frac{3}{2}\chi (H_u^+H_d^-+{\rm{c.c.}})\right)^2+\frac{1}{\alpha}({\rm{Im}}T)^2\\
\nonumber&+\frac{3}{2} \frac{\chi \lambda}{\sqrt{\alpha}}(S\bar{C}+{\rm{c.c.}})\left(\frac{1}{2} h^{2}+\left|H_{u}^{+}\right|^{2}+|H_d^-|^{2}\right)\\
\nonumber&+\frac{1}{ \alpha} |C|^{2}\left\{3+2\sqrt{6}\sigma +\frac{3}{2}\sigma^2+\frac{9}{4} \chi^{2}\left(\frac{1}{2} h^{2}+\left|H_{u}^{+}\right|^{2}+|H_d^-|^{2}\right)\right\}\\
\nonumber &+\frac{g^{2}+g^{\prime 2}}{8}\left(\frac{1}{2}h^{2} \cos 2 \beta+\left|H_{u}^{+}\right|^{2}-|H_d^-|^{2}\right)^{2}\\
&+\frac{g^2}{4} h^{2}\left\{\left|H_{u}^{+}\right|^{2}\sin^2 \beta+\left|H_{d}^{-}\right|^{2}\cos^2 \beta+\left(\frac{1}{2}H_u^+H_d^-\sin 2\beta e^{-i\gamma}+{\rm{c.c.}}\right)\right\},\label{full_VJ}
\end{align}
where $\gamma \equiv \delta_1+\delta_2$. Note that the scalar potential depend on the phases in the particular combination, $\gamma = \delta_1+\delta_2$,  while the other combination identified as the would-be neutral Goldstone boson does not appear in the potential. 

Then, we consider a minimization of the potential with respect to all fields other than $\sigma$ and $h$, which are treated as the background.\footnote{We assume that $\sigma$ and $h$ are slowly varying so that their time dependence are negligible.} 
From the expression~$\eqref{full_V}$ with Eq.~$\eqref{full_VJ}$, one can find that a point,
\begin{align}
H_u^+=H_d^-=S=C={\rm{Im}}T=\gamma=0, \ \ {\rm{and}}\ \   \beta=\pi/4, \label{SC}
\end{align}
satisfies a stationary condition. To see the stability of this extrema, we expand the fields around the extrema as
\begin{align}
\beta=\frac{\pi}{4}+\tilde{\beta},\ \   X=0+\tilde{X}, \ \ {\rm{with}}\ \ X=\{H_u^+,H_d^-,S,C,{\rm{Im}}T,\gamma \}   
\end{align}
up to quadratic order. The tilde on the fields denote the fluctuations. In the following, we discuss the stability of the fields individually.

\subsubsection*{Stabilization of $\beta$}
At the quadratic order, $\tilde{\beta}$ decouples from the other sectors, with the corresponding Lagrangian of the following form,
\begin{align}
- \frac{h^2}{2\Delta}(\partial_{\mu} \tilde{\beta})^2- \frac{1}{2}V_{\beta\beta}\tilde{\beta}^2,\label{beta}
\end{align}
where
\begin{align}
&V_{\beta\beta}=\left[-\frac{\lambda^2}{4}h^4-\frac{3\chi h^2}{4\alpha}\left(\sigma^2+\sqrt{6}\sigma+\frac{3\chi h^2}{4}\right)+\frac{g^{\prime 2}+g^2}{8}h^4\right]\frac{2}{\Delta^2},\\
&\Delta\equiv 1-\frac{h^2}{6}-\frac{\sigma^2}{6}.
\end{align}
Here, $V_{\beta\beta}$ is a second derivative of the potential with respect to $\beta$.

Substituting the explicit form of the background~$\eqref{sigma_app}$ and $\eqref{h_app}$,
a canonically normalized mass of $\beta$ is given by
\begin{align}
m_{\beta}^2 = \frac{3\frac{\lambda^2}{\alpha}+9(g^{\prime 2}+g^2)\frac{\xi}{\alpha}}{\lambda^2+36\frac{\xi^2}{\alpha}} = 4H^2\left(1+\frac{3\xi}{\lambda^2}(g^{\prime 2}+g^2)\right),
\end{align}
where we used Eq.~$\eqref{Inf_scale}$. Thus, since the $\beta$ direction gets a mass larger than the Hubble scale for $\xi (g^{\prime 2}+g^2)/\lambda^2\gg 1$, it is stabilized and decoupled during inflation.

\subsubsection*{Stabilization of charged Higgs}
The quadratic Lagrangian for the charged Higgs sector is summarized as 
\begin{align}
-\frac{1}{\Delta}|\partial_{\mu} \tilde{H}_{u}^{+}|^{2}-\frac{1}{\Delta}|\partial_{\mu} \tilde{H}_{d}^{-}|^{2}    -\left(\tilde{H}_{u}^{+*}, \tilde{H}_{d}^{-}\right)\left(\begin{array}{ll}
V_{++} & V_{+-} \\
V_{+-} & V_{--}
\end{array}\right)\left(\begin{array}{c}
\tilde{H}_{u}^{+} \\
\tilde{H}_{d}^{-*}
\end{array}\right),
\end{align}
where 
\begin{align}
&V_{++} =V_{--}=\left[\frac{2V}{3}\Delta+\frac{g^2}{8}h^2\right]\frac{1}{\Delta^2}, \\
&V_{+-}=\left[-\frac{\lambda^2}{4}h^2-\frac{3\chi }{4\alpha}\left(\sigma^2+\sqrt{6}\sigma+\frac{3\chi h^2}{4}\right)+\frac{g^2}{8}h^2\right]\frac{1}{\Delta^2},
\end{align}
and $V$ is the effective scalar potential during inflation, which is given by\footnote{This is exactly same as Eq.~$\eqref{V_h_sigma}$.}
\begin{align}
V=\left[\frac{\lambda^2}{16}h^4+\frac{1 }{4\alpha}\left(\sigma^2+\sqrt{6}\sigma+\frac{3\chi h^2}{4}\right)^2\right]\frac{1}{\Delta^2}.    
\end{align}
Taking into account a canonical normalization and diagonalizing the mass matrix, we find that the mass eigenvalues are given by
\begin{align}
&\left\{0,\ \  \frac{3\frac{\lambda^2}{\alpha}+9g^2\frac{\xi}{\alpha}}{\lambda^2+36\frac{\xi^2}{\alpha}}=4H^2\left(1+\frac{3g^2\xi}{\lambda^2}\right)\right\}.
\end{align}
The massless field is the would-be Goldstone boson eaten by the charged gauge boson.
We also find that the massive charged Higgs gets a mass of order the Hubble scale or beyond for $g^2\xi\lambda^2\gg 1$, so it is stabilized and decoupled safely during inflation.

\subsubsection*{Stabilization of $\gamma$ and ${\rm{Im}}T$}
Next, we investigate the decoupling of $\gamma$ and ${\rm{Im}}T (\equiv \tau)$, which also have a kinetic mixing in the following quadratic Lagrangian,
\begin{align}
-\frac{1}{2}\left(\partial_{\mu} \tilde{\gamma}, \partial_{\mu}\tilde{\tau}\right)\left(\begin{array}{ll}
a & c \\
c & b
\end{array}\right)\left(\begin{array}{l}
\partial^{\mu} \tilde{\gamma} \\
\partial^{\mu} \tilde{\tau}
\end{array}\right) -\frac{1}{2}V_{\gamma\gamma}\tilde{\gamma} ^2-\frac{1}{2}V_{\tau\tau}\tilde{\tau}^2,\label{gamma_tau}   
\end{align}
where 
\begin{align}
&a=\frac{h^2}{4\Delta}\left[1+\frac{h^2}{6\Delta}\left(1-\frac{3\chi}{2}\right)^2\right], \ \ b=\frac{2}{3\Delta^2},\ \ c=-\frac{h^2}{6\Delta^2}\left(1-\frac{3\chi}{2}\right),\\
&V_{\gamma\gamma}=-\frac{3\chi h^2}{8\alpha}\left(\sigma^2+\sqrt{6}\sigma+\frac{3\chi h^2}{4}\right)\frac{1}{\Delta^2}, \ \ V_{\tau\tau}=\frac{2}{\alpha \Delta^2}.
\end{align}
Since $ab-c^2=\frac{h^2}{6\Delta^3}>0$ and $a+b>0$, there is no ghost mode at the background. The kinetic matrix of Eq.~$\eqref{gamma_tau}$ can be canonically normalized by  
\begin{align}
\left(\begin{array}{c}
\tilde{\gamma} \\
\tilde{\tau}
\end{array}\right)=\frac{1}{\sqrt{2}}\left(\begin{array}{ll}
\left(\frac{\sqrt{\frac{b}{a}}}{\sqrt{ab}+c}\right)^{1/2} & -\left(\frac{\sqrt{\frac{b}{a}}}{\sqrt{ab}-c}\right)^{1/2} \\
\left(\frac{\sqrt{\frac{a}{b}}}{\sqrt{ab}+c}\right)^{1/2}  & \left(\frac{\sqrt{\frac{a}{b}}}{\sqrt{ab}-c}\right)^{1/2} 
\end{array}\right)\left(\begin{array}{c}
\tilde{\gamma}^{\prime} \\
\tilde{\tau}^{\prime}
\end{array}\right)\equiv \mathcal{M}\left(\begin{array}{c}
\tilde{\gamma}^{\prime} \\
\tilde{\tau}^{\prime}
\end{array}\right).
\end{align}
Note $\sqrt{ab}-c>0$. Then, we obtain
\begin{align}
-\frac{1}{2}\left(\partial_{\mu}\tilde{\gamma}^{\prime}\right)^{2}-\frac{1}{2}\left(\partial_{\mu}{\tilde{\tau}}^{\prime}\right)^{2} -\frac{1}{2}\left(\tilde{\gamma}^{\prime} , \tilde{\tau}^{\prime}\right)\mathcal{M}^{T}\left(\begin{array}{ll}
V_{\gamma\gamma} & 0 \\
0 & V_{\tau\tau}
\end{array}\right)\mathcal{M}  \left(\begin{array}{c}
\tilde{\gamma}^{\prime} \\
\tilde{\tau}^{\prime}
\end{array}\right) .
\end{align}
Then, after the diagonalization of the mass terms again, we obtain the mass eigenvalues as follows: 
\begin{align}
m^2_{\pm}&=  \frac{1}{2} \frac{V_{\gamma\gamma} b+V_{\tau\tau} a}{a b-c^{2}} \pm \frac{1}{2} \bigg[\bigg(\frac{c}{a b-c^{2}}\bigg)^{2}\bigg(V_{\gamma\gamma} \sqrt{\frac{b}{a}}+V_{\tau\tau} \sqrt{\frac{a}{b}}\bigg)^{2} \nonumber \\
&\qquad+\frac{1}{a b-c^{2}}\bigg(-V_{\gamma\gamma} \sqrt{\frac{b}{a}}+V_{\tau\tau} \sqrt{\frac{a}{b}}\bigg)^{2}\bigg]^{1/2}.  \label{mass_tau_gamma}
\end{align}
Substituting the background values~$\eqref{sigma_app}$ and $\eqref{h_app}$ into the above expressions, we make a further simplification of the results as
\begin{align}
&m^2_+= \frac{3+18\xi}{\alpha}=4(6\xi+1)\left(1+\frac{36\xi^2}{\alpha\lambda^2}\right)H^2,\\
&m^2_-=\frac{3 \lambda ^2}{\alpha  \lambda ^2+36 \xi ^2}=4H^2.
\end{align}
We find that both of mass eigenvalues in the $\gamma$ and ${\rm{Im}}T$ sector are larger than the Hubble scale.

\subsubsection*{Stabilization of $S$ and $C$}
Finally, we discuss the decoupling of the $S$-$C$ sector containing a mass mixing. The corresponding quadratic Lagrangian is given by
\begin{align}
-\frac{1}{\Delta}|\partial_{\mu}\tilde{S}|^2-\frac{1}{\Delta}|\partial_{\mu}\tilde{C}|^2-V_{S\bar{S}}|\tilde{S}|^2-\frac{1}{2}V_{SS}(\tilde{S}^2+{\rm{c.c.}})-V_{S\bar{C}}(\tilde{S}\tilde{C}^*+{\rm{c.c.}}) -V_{C\bar{C}}|\tilde{C}|^2,   
\end{align}
where
\begin{align}
&V_{S\bar{S}}= \left[\frac{\lambda^2}{2}h^2+\frac{2V}{3}\Delta\right]\frac{1}{\Delta^2},\ \ V_{SS}=-\frac{\lambda \rho}{2\Delta^2}h^2,\ \ V_{S\bar{C}}=\frac{3\chi \lambda }{4\sqrt{\alpha}\Delta^2}h^2,\label{V_ss}\\
&V_{C\bar{C}}=\left[\frac{1}{\alpha}\left(3+2\sqrt{6}\sigma+\frac{3}{2}\sigma^2+\frac{9\chi^2 h^2}{8}\right)+\frac{2V}{3}\Delta\right]\frac{1}{\Delta^2}.\label{V_cc}
\end{align}

Similarly to the previous cases, we introduce canonically normalized fields, $\tilde{S}'$ and $\tilde{C}'$, and divide them into real and imaginary components, as follows,
\begin{align}
\frac{1}{\sqrt{\Delta}}\tilde{S}=\tilde{S}'=\frac{1}{\sqrt{2}}({\rm{Re}}\tilde{S}'+i{\rm{Im}}\tilde{S}'), \ \ \frac{1}{\sqrt{\Delta}}\tilde{C}=\tilde{C}'=\frac{1}{\sqrt{2}}({\rm{Re}}\tilde{C}'+i{\rm{Im}}\tilde{C}').      
\end{align}
Then, in the above basis, we obtain the following mass matrices,
\begin{align}
-\frac{1}{2}\left({\rm{Re}}\tilde{S}^{\prime}, {\rm{Re}}\tilde{C}^{\prime}\right)\left(\begin{array}{ll}
V_{S\bar{S}}+V_{SS} & V_{S\bar{C}} \\
V_{S\bar{C}} & V_{C\bar{C}}
\end{array}\right)\Delta\left(\begin{array}{c}
{\rm{Re}}\tilde{S}^{\prime} \\
{\rm{Re}}\tilde{C}^{\prime}
\end{array}\right),\label{Re}
\end{align}
and
\begin{align}
-\frac{1}{2}\left({\rm{Im}}\tilde{S}^{\prime}, {\rm{Im}}\tilde{C}^{\prime}\right)\left(\begin{array}{ll}
V_{S\bar{S}}-V_{SS} & V_{S\bar{C}} \\
V_{S\bar{C}} & V_{C\bar{C}}
\end{array}\right)\Delta\left(\begin{array}{c}
{\rm{Im}}\tilde{S}^{\prime} \\
{\rm{Im}}\tilde{C}^{\prime}
\end{array}\right).\label{Im}
\end{align}

\begin{figure}[t]
 \begin{minipage}{0.5\hsize}
  \begin{center}
   \includegraphics[width=70mm]{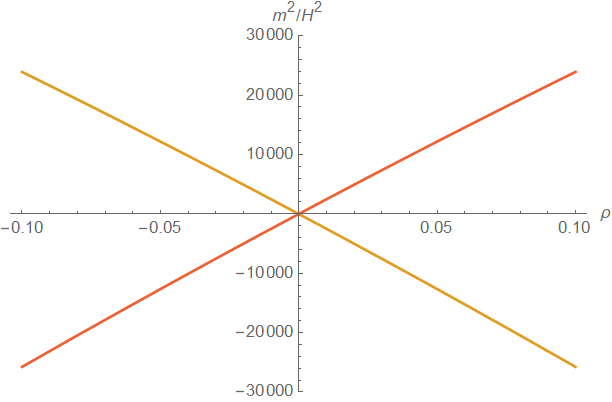}
  \end{center}
 \end{minipage}
 \begin{minipage}{0.5\hsize}
  \begin{center}
   \includegraphics[width=70mm]{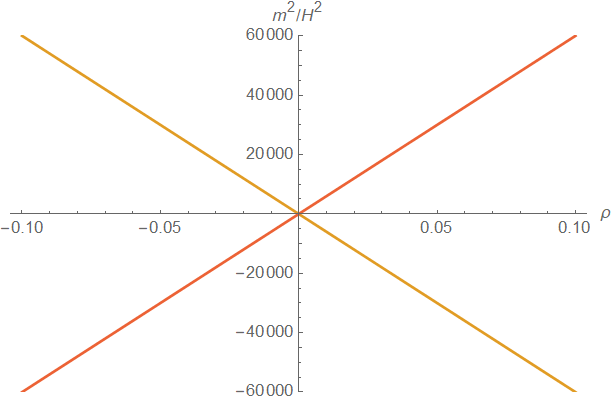}
  \end{center}
 \end{minipage}
  \caption{The squared mass eigenvalues for the lighter states of singlet scalars, $S$ and $C$, as a function of the singlet self-coupling $\rho$: $m^2_2$ (yellow) and $m^2_4$ (orange). We took the parameters as $(\lambda,\xi,\alpha)=(0.5,10^4,10^{10})$ for $R^2$-like inflation on left and  $(\lambda,\xi,\alpha)=(4\times 10^{-5},1,10^{2})$ for Higgs-like inflation on right.}
  \label{fig:SC_mass}
\end{figure}

Therefore, after diagonalizing the mass matrices in Eqs.~$\eqref{Re}$ and $\eqref{Im}$, we obtain the mass eigenvalues, respectively, as
\begin{align}
&m^2_{1,2}=\frac{\Delta}{2}\left[V_{S\bar{S}}+V_{SS}+V_{C\bar{C}} \pm \sqrt{\left(V_{S\bar{S}}+V_{SS}-V_{C\bar{C}}\right)^{2}+4 V_{S\bar{C}}^{2}}\right],\label{m12}
\end{align}
and
\begin{align}
m^2_{3,4}=\frac{\Delta}{2}\left[V_{S\bar{S}}-V_{SS}+V_{C\bar{C}} \pm \sqrt{\left(V_{S\bar{S}}-V_{SS}-V_{C\bar{C}}\right)^{2}+4 V_{S\bar{C}}^{2}}\right].\label{m34}
\end{align}
The above expressions are explicitly written down in the following,
\begin{align}
&m_{1,2}^2=\frac{18\xi\left( 6 \xi  (6 \xi +1)+\alpha  \lambda  (\lambda -\rho )\right)\pm 3f(\lambda,\alpha,\xi,\rho)}{2 \alpha  \left(\alpha  \lambda ^2+36 \xi ^2\right)},    \\
&m_{3,4}^2=m_{1,2}^2 \ \ {\rm{with}}\ \ \rho\rightarrow -\rho,
\end{align}
where
\begin{align}
f(\lambda,\alpha,\xi,\rho)  &=\Big[\alpha ^2 \lambda ^2 (6 \lambda  \xi +\lambda -6 \xi  \rho )^2+72 \alpha  \lambda  (6 \xi +1) \xi ^2 (6 \lambda  \xi +\lambda +6 \xi  \rho ) \nonumber \\
&\qquad+1296 (6 \xi +1)^2 \xi ^4\Big]^{1/2} . 
\end{align}

 In Fig.~\ref{fig:SC_mass}, we depict the behavior of the mass eigenvalues (normalized by $H^2$) in the $S$-$C$ sector. First, in two benchmark examples, one for $R^2$-like  inflation and  the other for Higgs-like inflation, we find that the heavier mass eigenvalues ($m^2_1$ and $m^2_3$) are always positive definite. However, we find that one of the lighter mass eigenvalues, namely, $m^2_2$ (yellow) and $m^2_4$ (orange), take negative values, independent of the parameter $\rho$, so there appears a tachyonic instability destabilizing the inflationary trajectory.

\begin{figure}[t]
 \begin{minipage}{0.5\hsize}
  \begin{center}
   \includegraphics[width=70mm]{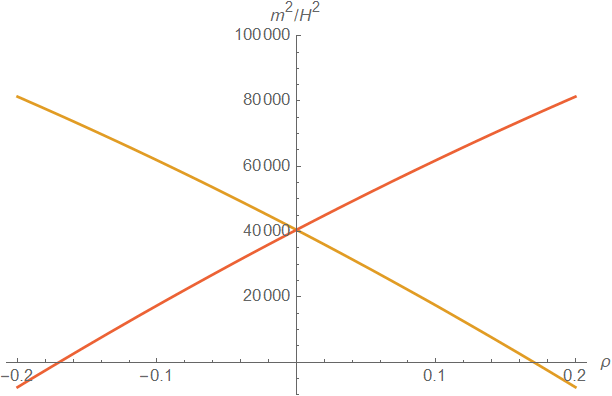}
  \end{center}
 \end{minipage}
 \begin{minipage}{0.5\hsize}
  \begin{center}
   \includegraphics[width=70mm]{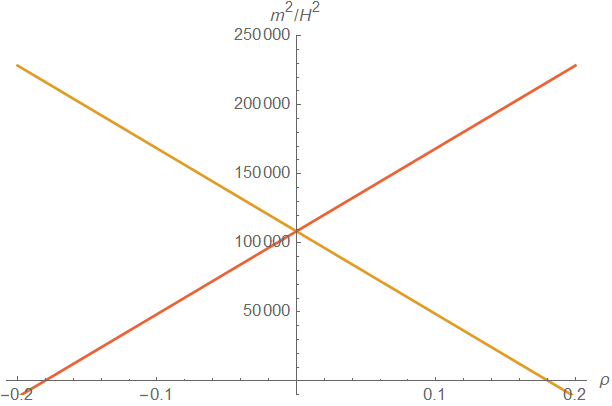}
  \end{center}
 \end{minipage}
    \caption{The same as Fig.~\ref{fig:SC_mass} but the quartic couplings in the frame function included. For both the $R^2$-like inflation on left and the Higgs-like inflation on right, we took $(\zeta_s,\zeta_c)=(3,0.4)$, and the canonical inflaton field value as $\phi=10$.}
  \label{fig:five}
\end{figure}

The aforementioned tachyonic mass problem is well known in the Higgs and Starobinsky inflation models in supergravity~\cite{Ferrara:2010yw,Ferrara:2010in,Kallosh:2013lkr}. One of the natural solutions for the tachyonic mass problem is to add quartic couplings for  $S$ and $C$ to the frame function, thus strongly stabilizing the tachyonic direction above the Hubble scale~\cite{Lee:2010hj,Ferrara:2010in,Kallosh:2013lkr}. Therefore,  the required higher order terms in the frame function are
\begin{align}
\Delta \Omega=-\zeta_s|S|^4-\zeta_c|C|^4\label{zeta_c}-\zeta_{sc}|S|^2|C|^2  
\end{align}
with $\zeta_s$, $\zeta_c$ and $\zeta_{sc}$ being real parameters. In particular, $-\zeta_c|C|^4$ corresponds to adding  
\begin{align}
-[\zeta_c\alpha^2 |X^0|^{-2}(\bar{\mathcal{R}}\mathcal{R})^2]_D, 
\end{align}
to Eq.~$\eqref{S_HD}$ in the dual picture. It was shown that the extra quartic couplings can be originated from the renormalizable couplings of $S$ or $C$ to vector-like heavy multiplets \cite{Lee:2010hj}.

The net effects of the extra quartic couplings in Eq.~(\ref{zeta_c}) are encoded in the modifications of $V_{S\bar{S}}$ and $V_{C\bar{C}}$ in Eqs.~$\eqref{V_ss}$ and $\eqref{V_cc}$, respectively, and $V_{S{\bar C}}$ and its complex conjugate, after the quadratic expansion of the potential with the extra couplings. As a result, the corrections to the mass terms for $S$ and $C$ are explicitly given in the following,\footnote{More general discussion on the mass corrections can be found in Refs.~\cite{Kallosh:2010ug,Kallosh:2010xz,Kallosh:2011qk}.}
\begin{align}
&\Delta V_{S\bar{S}}=\frac{\zeta_s\lambda^2}{4} \frac{h^4}{\Delta^2} \left(1+\frac{\sigma}{\sqrt{6}}\right)^{-2}, \\
&\Delta V_{C\bar{C}}=\frac{\zeta_c}{\alpha}\frac{1}{\Delta^2} \left(1+\frac{\sigma}{\sqrt{6}}\right)^{-2}\left(\sigma^2+\sqrt{6}\sigma+\frac{3\chi h^2}{4}\right)^2, \\
&\Delta V_{S\bar{C}}=\frac{\zeta_{sc}\lambda}{2\sqrt{\alpha}} \frac{1}{\Delta^2}\left(1+\frac{\sigma}{\sqrt{6}}\right)^{-2}h^2\left(\sigma^2+\sqrt{6}\sigma+\frac{3\chi h^2}{4}\right).
\end{align}

\begin{figure}[t]
 
 \begin{minipage}{0.5\hsize}
  \begin{center}
   \includegraphics[width=70mm]{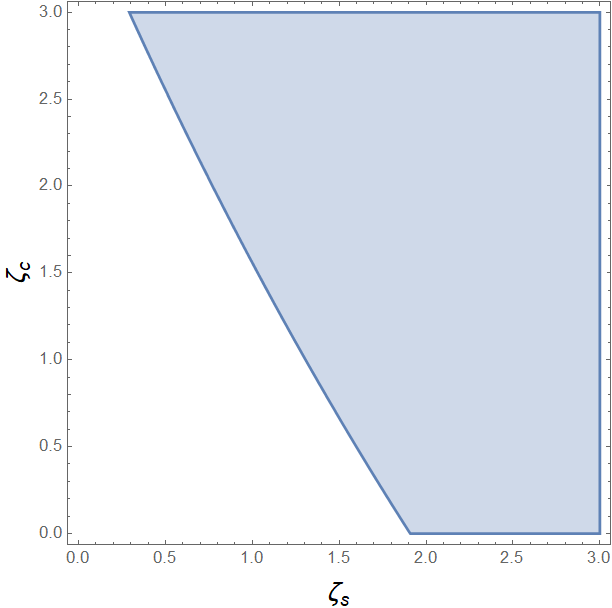}
  \end{center}
 \end{minipage}
 \begin{minipage}{0.5\hsize}
  \begin{center}
   \includegraphics[width=70mm]{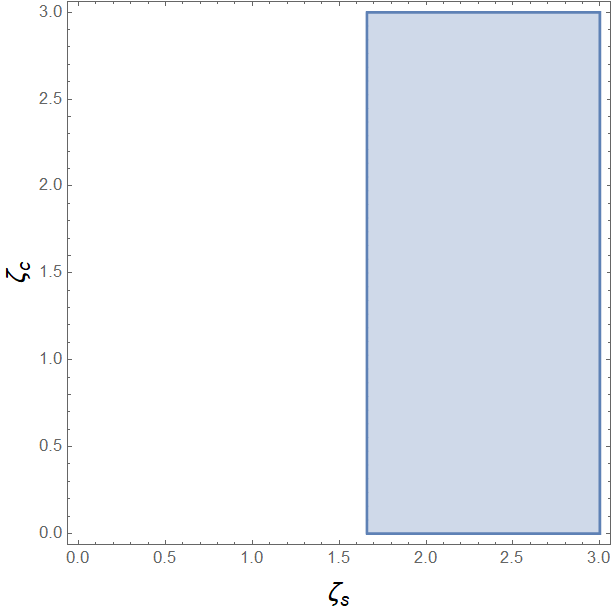}
  \end{center}
 \end{minipage}
    \caption{Parameter space for $\zeta_s$ and $\zeta_c$ satisfying $m_2>H$. We took  $(\lambda,\xi,\alpha)=(0.5,10^4,10^{10})$ for $R^2$-like inflation on left and  $(\lambda,\xi,\alpha)=(4\times 10^{-5},1,10^{2})$ on right for Higgs-like inflation. In both figures, we fixed $\rho=0.1$ and $\phi=10$.}
  \label{fig:six}
\end{figure}

As a result, we show  in Fig.~\ref{fig:five} that the effective squared masses for the lighter states of $S$ and $C$ ($m^2_2$ and $m^2_4$) in the presence of the quartic terms with $(\zeta_s,\zeta_c)=(3,0.4)$ and $\zeta_{sc}=0$. Thus, the otherwise tachyonic states of $S$ and $C$ can now take positive squared masses during inflation. Furthermore, in Fig.~\ref{fig:six}, we also show the parameter space for $\zeta_s$ and $\zeta_c$ satisfying $m_2>H$ for $\rho=0.1$ and $\zeta_{sc}=0$, and the canonical inflaton field value with $\phi=10$. We note that a large value of $\zeta_{sc}$ is undesirable because it tends to lower two of the eigenvalues ($m^2_{2}$ in Eq.~$\eqref{m12}$ and $m^2_{4}$ in Eq.~$\eqref{m34}$). Therefore, we can conclude that it is sufficient to introduce a nonzero $\xi_s$ in the frame function for the stability of the $S$-$C$ sector.  However, as will be discussed in the next section, a nonzero $\zeta_c$ is desirable for the SUSY breakdown in the local vacuum with a vanishing small cosmological constant.

\section{SUSY breaking from dual superfields}\label{Pheno}
We consider some phenomenological implications of the model in the low energy after the inflation. In the minimal setup for inflation in our model, the VEVs for sigma and Higgs fields vanish in the vacuum, so SUSY would be unbroken in our model. Thus, we introduce several extensions of the model for supersymmetry breaking and its transmission to the NMSSM, and discuss the effects of the hidden sector fields on them.

\subsection{Higher curvature terms for SUSY breaking}

As a possible extension of our model without introducing an extra hidden sector for SUSY breaking, we introduce extra curvature terms~\cite{Dalianis:2014aya}, as follows,
\begin{align}
S=[|X^0|^2f(\mathcal{R}/X^0,\bar{\mathcal{R}}/\bar{X}^{\bar{0}})]_D ,  
\end{align}
where
\begin{align}
f=-3+  \alpha|\mathcal{R}/X^0|^2 -\gamma_c \alpha \left(\mathcal{R}/X^0+{\rm{c.c.}}\right)-\zeta_c\alpha^2|\mathcal{R}/X^0|^4. \label{def_f}
\end{align}
We note that the effects of $\zeta_c$  for the stability of inflation were discussed already introduced in Eq.~$\eqref{zeta_c}$, although it is not necessary. But, for SUSY breaking, we need not only $\zeta_c$  but also a linear term  $\gamma_c$ in $\mathcal{R}$. In general, a function of the curvature multiplet $\mathcal{R}$ does not produce higher order terms of the Ricci scalar $R^k$ with $k\geq 3$ in components, so the Starobinsky structure with $\sim R^2$ is preserved.

In the dual description for higher curvature terms, we have the modified frame function and the unmodified superpotential in the $C$ and $T$ sector as follows,
\bea
 \Omega &=&-3+(T+{\bar T})+|C|^2-\gamma_c (C+\bar{C})-\zeta_c |C|^4,\label{dual_2} \\
W&=&\frac{1}{\sqrt{\alpha}}\, TC.
\eea
Here, we  performed the same duality transformation as discussed in Sec.~\ref{sec_dual}. 

From the scalar potential for $T$ and $C$ superfields, we can identify the local Minkowski minimum with SUSY breakdown \cite{Dalianis:2014aya} at 
\bea
\langle C \rangle &=&\frac{1}{54\zeta_c} \,\Big(1+\sqrt{1+324 \zeta_c} \Big)\equiv c_0, \\ 
\langle T \rangle &=& \gamma_c c_0 + c^2_0 (1-6\zeta_c c^2_0)\equiv t_0,
\eea
subject to the conditions,
\bea
\gamma_c=-c_0 + \frac{2}{c_0} \bigg(1+\frac{1}{3} c^2_0\bigg)
\eea
and $9-36 \zeta_c c^2_0>0$.  We note that the latter condition is satisfied for $\zeta_c<0.48$ while we need $\zeta_c>0.15$ for $c_0<1$.  In this case, the linear coupling in the frame function is constrained to $1.7<\gamma_c<2.5$.

As a result, we obtain nonzero $F$-terms for $C$ and $T$ superfields by
\bea
F^C &=& -e^{K/2} K^{C{\bar C}} (D_C W)^\dagger - e^{K/2} K^{C{\bar T}} (D_T W)^\dagger, \\
F^T &=&  -e^{K/2} K^{T{\bar T}} (D_T W)^\dagger - e^{K/2} K^{T{\bar C}} (D_C W)^\dagger,
\eea
with
\bea
D_C W &=& \frac{1}{9\sqrt{\alpha} }\, (3+c^2_0) \bigg(\frac{246-192c_0+55c^2_0+27 c^3_0}{66-5c^2_0}\bigg), \\
D_T W &=&  \frac{1}{9\sqrt{\alpha} }\, c_0\bigg(\frac{39-14c^2_0}{66-5c^2_0}\bigg),
\eea
and the gravitino mass is given by
\bea
m_{3/2}= \frac{243}{8}\sqrt{\frac{6}{\alpha}}\,\cdot \frac{c_0(3+c^2_0)^{1/2}}{(66-5c^2_0)^{3/2}}. 
\eea
Here, $c_0$ is constrained by $c_0<\sqrt{66/5}$. Therefore, we find that the F-terms are of order $F^C\sim F^T\sim M_P m_{3/2}$ and the gravitino mass is $m_{3/2}\sim M_P/\sqrt{\alpha}$. Since perturbativity constrains $\alpha\lesssim 10^{10}$,  the gravitino mass is given by $m_{3/2}\gtrsim 10^{13}\,{\rm GeV}$, so  a high-scale SUSY breaking is favored.

\subsection{O'Raifeartaigh model for SUSY breaking}

Instead of higher curvature terms, we consider an alternative possibility for SUSY breaking where another singlet chiral superfield $\Phi$ is introduced, with the following frame function and the renormalizable superpotential of O'Raifeartaigh type, as follows,
\bea
\Omega&=& -3-(T+{\bar T})+|C|^2+|\Phi|^2 -\gamma\, |\Phi|^4, \label{framephi} \\
W &=& \frac{1}{\sqrt{\alpha}}\, TC + \kappa\, \Phi + g\,  \Phi C^2 +\lambda \Phi^3 +\kappa' C + g' \Phi^2 C + \lambda' C^3 \label{Osuper}
\eea
where $\kappa, g,\lambda, \kappa', g', \lambda'$ are the extra parameters in the superpotential and $\gamma$ is the quartic coupling of $\Phi$ in the frame function.
Here, we remark that the $Z_{4R}$ R-symmetry can ensure the above form of the superpotential, with $R$-charge assignments, $R[\Phi]=R[C]=+2$ and $R[T]=0$. We note that the dual scalar superfield $T$ is neutral under the $Z_{4R}$ R-symmetry, because the corresponding frame function takes the form of $T+{\bar T}$. 

First, for simplicity, as in the standard O'Raifeartaigh model, we set $\lambda=\kappa'=\lambda'=g'=0$.
Then, there is a minimum with $C=T=0$, for which we obtain a nonzero F-term, $F_\Phi=\kappa$, and $F_C=F_T=0$. 
We note that $\Phi$ would be a pseudo-flat direction for $\gamma=0$ in the frame function in eq.~(\ref{framephi}), but it can be stabilized by loop corrections \cite{flat} or higher order terms \cite{Kitano:2006wz} in the K\"ahler potential.  That is, for $\gamma\neq 0$, we obtain the squared mass for $\Phi$ as $m^2_\Phi=4\gamma\kappa^2/M^2_P$.
In this model, the coupling between $\Phi$ and  $C$ gives rise to a mass splitting in the $T$ and $C$ sector, with mass eigenvalues for scalars and the fermion mass, respectively, given by
\bea
m^2_{s,\pm} &=& \frac{M^2_P}{\alpha} \pm 2g\kappa, \label{simplescalar} \\
m_f &=&  \frac{M_P}{\sqrt{\alpha}}. 
\eea
As a result, the SUSY breaking effects are controlled by $\kappa$, so it is possible to get a low-scale SUSY breaking for an appropriate choice of $\kappa$ in this case.  We also note that the condition for a vanishing cosmological constant in supergravity gives rise to the gravitino mass as $m_{3/2}=|F_\Phi|/(\sqrt{3}M_P)=\kappa/(\sqrt{3}M_P)$.

We also comment on the effects of the general couplings in the superpotential in eq.~(\ref{Osuper}).
In this case, the local minimum is shifted to $C=0$, $T=-\sqrt{\alpha}\,\kappa'$, due to a nonzero $\kappa'$, and the would-be pseudo-flat direction can be still stabilized at $\Phi=0$ due to the extra quartic term for $\Phi$ in the frame function. Then, we obtain $F_C=F_T=0$ and $F_\Phi=\kappa$, as in the case with $\lambda=\kappa'=\lambda'=g'=0$.
Taking the couplings in the superpotential in eq.~(\ref{Osuper}) to be real and including a nonzero quartic correction for $\Phi$ in the frame function, we find that the squared mass matrices for $({\rm Re}\,C,{\rm Re}\,\Phi)$ and $({\rm Im}\,C,{\rm Im}\,\Phi)$ are given, respectively, by
\bea
M^2_R =\left(\begin{array}{cc} \frac{M^2_P}{\alpha}+2g \kappa  & 2g' \kappa \\ 2g' \kappa & 6\lambda \kappa+m^2_\Phi \end{array}\right),  \quad 
M^2_I =\left(\begin{array}{cc} \frac{M^2_P}{\alpha}-2g \kappa & -2g' \kappa  \\ -2g' \kappa & -6\lambda \kappa+m^2_\Phi \end{array}\right)
\eea 
with
\bea
m^2_\Phi\equiv \frac{4\gamma \kappa^2}{M^2_P}.
\eea
Here, we note that for $g'=\lambda=0$ and $\gamma=0$, the former result in eq.~(\ref{simplescalar}) is recovered, namely, the $C$ field is stabilized at $C=0$, and the $\Phi$ field becomes a flat direction. 
But, for general extra couplings, we obtain the mass eigenvalues for $({\rm Re}\,C,{\rm Re}\,\Phi)$ and $({\rm Im}\,C,{\rm Im}\,\Phi)$, respectively, as
\bea
m^2_{s1,s2} &=&\frac{1}{2} \bigg[\frac{M^2_P}{\alpha}+2(g+3\lambda)\kappa+m^2_\Phi \pm \sqrt{\Big(\frac{M^2_P}{\alpha}+2(g-3\lambda)\kappa-m^2_\Phi \Big)^2+16g^{\prime 2}\kappa^2} \bigg], \\
m^2_{s3,s4} &=&\frac{1}{2} \bigg[\frac{M^2_P}{\alpha}-2(g+3\lambda)\kappa+m^2_\Phi  \pm \sqrt{\Big(\frac{M^2_P}{\alpha}-2(g-3\lambda)\kappa-m^2_\Phi \Big)^2+16g^{\prime 2}\kappa^2} \bigg].
\eea
Therefore, as far as the following conditions are satisfied,
\bea
 \frac{M^2_P}{\alpha}>2g \kappa, \quad m^2_\Phi > 6\lambda \kappa, \label{stability1}
\eea
and
\bea
\bigg(\frac{M^2_P}{\alpha}-2g \kappa\bigg)\Big(m^2_\Phi-6\lambda \kappa  \Big)> 4g^{\prime 2} \kappa^2,
\eea
 $m^2_{s3,s4}$ are positive definite, and $m^2_{s1,s2}$ are also necessarily positive. 
In particular, the second condition in  eq.~(\ref{stability1}) corresponds to
\bea
\frac{\gamma}{\lambda}> \frac{3M^2_P}{2\kappa}.
\eea
In this case, there appear a stable local minimum for SUSY breaking, which is still controlled by $\kappa$, as far as the above stability condition for $\kappa$ and the dimensionless parameters, $\gamma$ and $\lambda$, is satisfied.

\subsection{Comments on soft masses in the visible sector}

In the presence of a non-minimal coupling to the Higgs fields in NMSSM and the dual superfields from $R^2$-supergravity, there are extra contributions to the $\mu$ term proportional to the gravitino mass \cite{Lee:2010hj}, in addition to the tree level contribution in NMSSM, as follows,
\bea
\mu=\lambda\langle {\tilde S}\rangle + \frac{3}{2}\,\chi m_{3/2} -\frac{1}{2}\chi K_{\bar I} {\bar F}^{\bar I}. \label{muterm}
\eea
Here, we have rescaled the superfields in the NMSSM Higgs sector in Einstein frame by ${\tilde H}_{u,d}=e^{{\cal K}/6} H_{u,d}$ and ${\tilde S}=e^{{\cal K}/6} S$, etc. The second term  is due to the non-minimal coupling $\chi$ as found in Jordan frame supergravity in Ref.~\cite{Lee:2010hj} and the third term is a new Giudice-Masiero contribution \cite{GM}, coming from the contact interactions between the dual superfields and the Higgs superfields in the K\"ahler potential, ${\cal K}\supset \frac{3}{2}\chi Z H_u H_d$ with $Z=e^{{\cal K}_0/3}$ where ${\cal K}_0$ is the  part of the K\"ahler potential containing the SUSY breaking fields, such as $C, T$ or $\Phi$. Thus, not only the VEV of the NMSSM singlet $S$ but also the gravitino mass and the SUSY breaking scale determine the $\mu$ term at a naturally small value.  For a large $\chi$, the $\mu$ term is typically much larger than the gravitino mass \cite{Lee:2010hj}.

Next we remark on the transmission of SUSY breaking to the visible sector that is applicable to both mechanisms for SUSY breaking via higher curvature terms or an O'Raifeartaigh model. 
In our construction for Jordan frame supergravity, the visible sector and the hidden sector composed of $T, C$ and $\Phi$ are sequestered in the frame function \cite{sequester}, so soft masses in NMSSM vanish at tree level. 
However, anomaly mediation is always present \cite{sequester,anomalymed}, so the soft masses in the visible sector are at least loop-suppressed as compared to gravitino mass.  In order to cure the problem with tachyonic slepton masses, we can introduce gravity mediation by adding the contact terms between the visible sector and the hidden sector in the frame function,
\bea
\Omega_{\rm contact} = C_{{\bar\alpha}{\beta}} X^\dagger X z^\dagger_{\bar \alpha}  z_\beta +{\rm c.c}
\eea
where $X=C,\Phi$, and $z_\alpha$ are NMSSM superfields, and $ C_{{\bar\alpha}{\beta}} $ are the coupling parameters.
However, for $C_{{\bar\alpha}{\beta}}\neq \delta_{{\bar\alpha}{\beta}}$, gravity mediation would cause dangerous flavor problems \cite{falkowski}.
Instead, we can consider alternative mediation mechanisms such as gauge mediation, $U(1)'$ mediation, etc.

\section{Conclusions}\label{summary}

We proposed a new supergravity construction for the Higgs-$R^2$ inflation as a UV completion of the Higgs inflation. 
A nontrivial Higgs potential during inflation requires the NMSSM extension in the visible sector, whereas the supersymmetric $R^2$ term gives rise to dual chiral superfields, $T$ and $C$, in the dual-scalar supergravity. We introduced equivalent frames (namely, Jordan frame, Einstein frame and sigma model frame) for the supergravity Lagrangian in the superconformal framework, among which the sigma model frame makes the conformal symmetry and the validity of unitarity up to the Planck scale more manifest.  

We have shown that the slow-roll inflation can be realized from the mixture of the SM Higgs and the real part of $T$ (scalaron or sigma field), and ensured the stability of the slow-roll inflation from the decoupling conditions for extra scalar fields in the model. Then, we found that not only all the MSSM scalars but also the directions of the NMSSM singlet scalar $S$ and the spectator dual scalar $C$ are decoupled during inflation, at the expense of introducing the extra quartic coupling  for $S$  in the frame function. 

As low-energy remnants of the Higgs-$R^2$ supergravity, we have  suggested the possibilities for SUSY breaking in the vacuum, in the presence of either modified higher curvature terms or an extra singlet  chiral superfield of O'Raifeartaigh type.
We found that the non-minimal coupling to the Higgs fields and the couplings between the dual superfields  and the Higgs fields in the frame function give rise to naturally small contributions to the $\mu$ term after SUSY is broken. 
We also pointed out that soft SUSY breaking terms in the visible sector vanish at tree level due to the sequestered form of the frame function in Jordan frame supergravity, but they can be generated by anomaly mediation, subject to generically flavor-violating gravity mediation as well as other mediation mechanisms.


\subsection*{Acknowledgements}
The work is supported in part by Basic Science Research Program through the National Research Foundation of Korea (NRF) funded by the Ministry of Education, Science and Technology (NRF-2019R1A2C2003738 and NRF-2021R1A4A2001897). 
The work of AGM is supported in part by the Chung-Ang University Young Scientist Scholarship in 2019.

\end{document}